\documentclass[lettersize,journal]{IEEEtran}
\usepackage{amsmath,amsfonts}
\usepackage{algorithmic}
\usepackage{algorithm}
\usepackage{array}
\usepackage[caption=false,font=normalsize,labelfont=sf,textfont=sf]{subfig}
\usepackage{textcomp}
\usepackage{stfloats}
\usepackage{url}
\usepackage{verbatim}
\usepackage{graphicx}
\usepackage{cite}
\usepackage{lineno,hyperref}
\modulolinenumbers[5]
\usepackage{algorithmic}
\usepackage[dvipsnames]{xcolor}
\usepackage{mathtools}
\usepackage{enumerate}
\usepackage{booktabs}
\usepackage{color}
\newcommand{\mat}[1]{\boldsymbol{#1}}

\newcommand{\norm}[1]{\left\lVert#1\right\rVert}
\newcommand{\R}{\ensuremath{\mathbb{R}}}

\begin{document}

\title{A cross-domain recommender system using deep coupled autoencoders}


\author{Alexandros Gkillas,  ~\IEEEmembership{Member,~IEEE}
, Dimitrios Kosmopoulos, ~\IEEEmembership{Member,~IEEE}
\thanks{This work has been submitted to the IEEE for possible publication. Copyright may be transferred without notice, after which this version may no longer be accessible.\\
A. Gkillas is with Graduate department of Electrical and Computer Engineering, Patras University, Patras,
Greece}
\thanks{D.Kosmopoulos is with the Faculty of
Department of Computer Engineering and Informatics,  Patras University, Patras,
Greece.}
}
\makeatletter
\def\ps@IEEEtitlepagestyle{%
  \def\@oddfoot{\mycopyrightnotice}%
  \def\@oddhead{\hbox{}\@IEEEheaderstyle\leftmark\hfil\thepage}\relax
  \def\@evenhead{\@IEEEheaderstyle\thepage\hfil\leftmark\hbox{}}\relax
  \def\@evenfoot{}%
}
\def\mycopyrightnotice{%
  \begin{minipage}{\textwidth}
  \centering \scriptsize
  Copyright~\copyright~20xx IEEE. Personal use of this material is permitted. Permission from IEEE must be obtained for all other uses, in any current or future media, including\\reprinting/republishing this material for advertising or promotional purposes, creating new collective works, for resale or redistribution to servers or lists, or reuse of any copyrighted component of this work in other works by sending a request to pubs-permissions@ieee.org.
  \end{minipage}
}
\makeatother





\maketitle

\begin{abstract}
Long-standing data sparsity and cold-start constitute thorny and perplexing problems for the recommendation systems. Cross-domain recommendation as a domain adaptation framework has been utilized to effectively address these challenging issues, by exploiting information from multiple domains. In this study, an item-level relevance cross-domain recommendation task is explored, where two related domains, that is, the source and the target domain contain  common items. Additionally, a user-level relevance scenario is considered, where the two related domains contain common users.  In light of these scenarios, two novel coupled autoencoder-based deep learning methods are proposed for cross-domain recommendation. The first method aims to simultaneously learn a pair of autoencoders in order to reveal the intrinsic representations in the source and target domains, along with a coupled mapping function to model the non-linear relationships between these representations. The second method is derived based on a new joint regularized optimization problem, which employs two autoencoders to generate in a deep and non-linear manner the user and item-latent factors, while at the same time a data-driven function is learnt to map the latent factors across domains. Extensive numerical experiments  are conducted illustrating the superior performance of our proposed methods compared to several state-of-the-art cross-domain recommendation frameworks. 
\end{abstract}


\section{Introduction}
\IEEEPARstart{R}{ecommender} systems are automated applications that suggest products to consumers based on their observed interests \cite{RCR,10.1016/j.neucom.2016.12.102,Ma2019}. A user's preferences in items is stored in the form of interaction, such as numerical rating, within a rating matrix. As a result, users, items, and the rating matrix form a domain \cite{Bobadilla2013}. The issues of cold start, sparsity and inclusion of new customers or products may compromise the performance of recommenders \cite{cold,Natarajan2020,Zhang20201}. While these problems have been extensively studied from a single-domain viewpoint, cross-domain recommender systems (CDRS) bring a different perspective to their solution \cite{cross_1}.

The challenge of recommending specific items to consumers in a target domain (e.g., a resource-scarce market) by using data from neighboring high-resource domains, e.g., using data from a much larger market to boost recommendations in a target market, is central to the principle of cross-domain recommendation \cite{app1,app2,app3,app4,8392508}. We hypothesize that data from one domain can be used to boost advice in a different domain. Such an approach has attracted the interest of many researchers in the recent years, e.g.,  
\cite{pmlr-v38-iwata15,Cantador2015,Zhu2021,Zhong2020,Moreno2012,text1,text2,Gao2019,He2019}. 

{The cross-domain recommendation problem (CDR) has been explored under different perspectives and scenarios, aiming  to address the challenging recommendation issues,
that is, the data sparsity  and the cold start problems. Particularly, the  sparsity issue arises due to user interactions with a small portion of items in the particular  domain, whereas the cold start problem arises due to the lack of data about new entities, i.e., a new item/new user \cite{Natarajan2020}}. 
In general, CDR methods can be divided into three major categories, that is content-based frameworks, embedding-based frameworks and rating pattern-based approaches \cite{Zhu2021}. \textit{Content-based} approaches examine the CDR problem from a content-level relevance point of view. Particularly,  this type of methods aim to link various domains by capturing and utilizing similar content information, such as, user-generated reviews \cite{text1}. Contrary to these methodologies, \textit{embedding-based} approaches explore the CDR problem from a user-relevance or item-relevance perspective. Exploiting common users and/or common items, this category extracts embedding knowledge (e.g., user/item latent factors) and then transfer it  across domains through domain adaptation techniques such as neural networks \cite{10.1145/3366423.3380036}, \cite{man} and transfer learning \cite{zhang,T1,T2}. Finally, \textit{rating pattern-based} approaches aim to transfer  information such as rating patterns from the source to the target domain \cite{BLin,Loni,Darec}. 

The proposed methods belong to the category of embedding-based approaches. 
{In more detail, we explore two   cross-domain recommendation tasks. In the first task, we examine an \textbf{item-level relevance} recommendation task  assuming that the two related domains, that is, the source and the target domain, contain  common items  (item full overlap) with different
users (user non-overlap) without sharing any  information regarding the users' behavior (ratings), and thus avoiding the leak of user privacy (See Figure \ref{fig:items_scenario}).} Moreover, in the second task, we consider a \textbf{user-level relevance} task  where the two related domains contain  common users  (user full overlap) with different
items (item non-overlap) (See Figure \ref{fig:users_scenario}).
Our contribution concerns two novel coupled autoencoder-based deep learning methods for cross-domain recommendation:
\begin{itemize}
    \item The first (more efficient) method, dubbed CACDR (Coupled Autoencoder Cross - Domain Recommendation) aims to simultaneously learn a pair of autoencoders in order to reveal the intrinsic representations of the common entities (common items or users) in the source and target domains along with a coupled mapping function to model the non-linear relationships between these representations; thus it is able to transfer beneficial information from the source to the target domain.
    \item The second method, dubbed LFACDR (Latent Factor Autoencoder Cross-Domain Recommendation) is derived based on a new joint regularized optimization problem, which employs two autoencoders to generate in a deep and non-linear manner the user and item-latent factors; at the same time a data-driven function is learnt to map the item-latent factors (item-level relevance scenario) or user-latent factors (user-level relevance scenario) across domains. The LFACDR is more computationally demanding compared to the CACDR, but offers better performance.
\end{itemize}
   Different from other studies we optimize the autoencoders jointly, thus learning in an end-to-end fashion the intrinsic relationships across domains. 

The rest of the paper is organized as follows: section \ref{sec:related} gives an overview of the research related to ours; section \ref{sec:problem_form} states the problem formulation;section \ref{sec:models} analyzes the proposed methods; section \ref{sec:exp} validates experimentally the proposed methods using two public datasets as source and target domains; finally, section \ref{sec:conc} summarizes our contributions and gives future directions.

\section{Related Works}
\label{sec:related}

In literature, there is a plethora of studies attempting to address the challenging recommendation issues that emerge, that is, the data sparsity and the cold start by developing CDR strategies. In recent years, the problem of CDR has been tackled from multiple perspectives and different assumptions, thus rendering this problem particularly difficult to describe  under a unique generic framework \cite{Cantador2015}, \cite{Zhu2021}. To that end, in this section several representative studies are briefly presented.

Specifically, Singh et al. \cite{Singh2008} used a matrix factorization approach to transfer information across domains by sharing the user latent factor. Pan et al. \cite{Pan2010} employed a principled matrix-based transfer learning methodology to extract and transfer knowledge concerning the users and items from the source to the target domain. Agarwal, et al. \cite{lfm} proposed a collective matrix factorization method exploiting correlated information  across domains via localized factor models to tackle the sparsity problem in the target domain. Moreno et al. \cite{Moreno2012} exploited the information from multiple domains in order to improve the recommendation accuracy for the target domain. Lian et al. \cite{CCCFNet} combined collaborative filtering and content-based filtering into a multi-view neural network to tackle the CDR problem. Aiming to overcome the data sparsity problem other studies such as \cite{cluster1}, \cite{cluster2}, \cite{rafa}, \cite{cluster3} employed cluster-level matrix factorization techniques to share common information between users and items across domains. 

Recently, Man et al. \cite{man}  used a matrix factorization method under the user sharing assumption to extract the latent factors models and a multi-layer perceptron to model and transfer valuable knowledge across domains. Kang et al. proposed a semi-supervised mapping to recommend items for cold start users by exploiting the distribution of shared users across domains. Elkahky et al. \cite{Elkahky} proposed a deep learning methodology to map shared users and items to a hidden space where the similarity between users and items is maximized. 
Zhong et al. \cite{Zhong2020} utilized a deep learning architecture based on the autoencoders and an attention mechanism to extract and fuse information from multiple closely-related domains, thus enhancing the rating prediction accuracy. Zhu et al. \cite{At} proposed a graphical and attentional framework for the CDR problem based on the rating and content information across domains.
Additionally, Kanagawa et al. \cite{Kanagawa} employed an unsupervised domain adaptation approach by reformulating the recommendations as an extreme classification task. Zhao et al. \cite{Zhao1}  captured the interactions of different domains as a whole, and propagated user preferences, based on graph neural networks. 
He et al. \cite{He2019} proposed a codebook transfer learning procedure to learn the proper codebook scale balancing both the computational complexity and prediction accuracy for CDR. Gao et al. \cite{Gao2019} examined the CDR from a data privacy perspective without sharing any information about the users' data. Ma et al. \cite{Ma2021} addressed the problem of insufficient common users by employing a fully connected trust-aware deep learning framework to discover the intrinsic relationships between common and non-common users. Iwata et al. \cite{pmlr-v38-iwata15} proposed a CDR architecture assuming that the user and item-latent factor models in different domains are derived from a common Gaussian distribution. Hu et al.\cite{Conet} used deep cross connection networks to exploit and transfer information across the domains. Li et al. \cite{Li2020} exploited the merits of the dual transfer learning and the latent embedding methodology to tackle the CDR problem. In more details, an orthogonal matrix was employed to transfer the knowledge from the source to target domain. Yuan et al. \cite{Darec} utilized a deep domain adaption model to extract and transfer patterns from rating matrices in different domains, without considering any auxiliary information. Furthermore,  
{Zhang et al. \cite{9640532} proposed a dual adversarial network, thus forming a two-way recommendation model to enhance the recommendation for both the source and target domain. 
Wang et al. \cite{8698453} addressed the issues of data sparsity and data imbalance in the cross-domain recommendation problem   by employing an adversarial and transfer learning procedure to transfer information from different domains to the target domain. Zhang et al. \cite{8525418} utilized a cross-domain recommendation kernel-induced knowledge transfer approach considering partially overlapped entities.}

To proceed further, other studies focused on extracting and utilizing content information to tackle the CDR problem.
Xin et al. \cite{text1} proposed a CDR framework utilizing review text to alleviate data sparsity limitations. Along the lines of the previous method, Fu et al. \cite{text2} utilized stacked denoising autoencoders to fuse review text with the rating matrices to tackle the data sparsity and cold start problems in the target domain. Zhao et al. \cite{ctan} extracted multiple aspects of users and items based on review documents aiming to learn aspect correlations across domains via an attention mechanism. 

Similar to our work, the study in \cite{man} belongs to the embedding-based frameworks considering also an item-level relevance cross-domain recommendation task. However, this approach employs matrix factorization frameworks to extract the user and item-latent factors, thus rendering it limited only to capture  linear and rather shallow features from the complex and non-linear collaborative relationships of the  users and items. Furthermore, taking into consideration that the CDR problem is a domain adaptation procedure the learning of the latent factors of the source and target domains independently may result in poor performance, since there is no influence or transferred knowledge between the domains during the learning stage.

Different from the above-mentioned approach, in our study the goal is to capture  and model the underlying relationships between the users and items from the source and target domain in a deep and non-linear manner; we employ two novel cross-domain recommendation frameworks based on the coupled autoencoders. We optimize jointly the autoencoders of the source and the target domain transferring valuable information across domains during the training stage. Autoencoders have also been employed in other studies addressing different cross-domain recommendation scenarios, such as multi-domain recommendation tasks \cite{Zhong2020}, \cite{Li2020}, \cite{Darec} and content-based recommendation tasks \cite{text1}. \emph{Nevertheless, in these methods the autoencoders of the source and target domains are trained independently. This procedure is piecemeal and thus sub-optimal, since there is no transfer or coupled learning  the source domain  and the target domain and hence no influence from one to another during the training process.} Contrary to these learning procedures, in this study we argue that better and more meaningful intrinsic representations  can be derived, not only based on the available input data in each domain separately, but also taking into account the internal relationships that exist across domains during the learning of the autoencoders.

\section{Notations and Problem Formulation} \label{sec:problem_form}
Table \ref{tab:notation} summarizes all the required notations of this study. In the literature the two domains are often referred as the 'source' and 'target' domains.  Without loss of generality, let $\mat{R}^s \in \R^{m \times n}$ and $\mat{R}^t \in \R^{m \times n}$ be the rating matrices representing the ratings between $m$ items and $n$ users for the source and the target domain respectively, where $\mat{R}^s(i,j)$ represents the rating of the user $j$ for the item $i$ in the source domain and $\mat{R}^t(i,j)$ represents the corresponding rating in the target domain. Furthermore, we denote as $\mat{M}^s=[[\mat{m}_1^s]^T; [\mat{m}_2^s]^T; ...] = \mat{R}^S$, $\mat{M}^t=[[\mat{m}_1^t]^T; [\mat{m}_2^t]^T; ... ]=\mat{R}^t$ the item rating matrix of the source and target domain and $\mat{U}^s=[[\mat{u}_1^s]^T; [\mat{u}_2^s]^T; ...] = \mat{{R}^s}^T$, $\mat{U}^t=[[\mat{u}_1^t]^T; [\mat{u}_2^t]^T; ... ]=\mat{{R}^t}^T$ the user rating matrix of the source and the target domain, respectively. In general, the item rating vector $\mat{m}_i^s \in \R^{1 \times n}$ describes the rating relationship between the item $i$ and all the users in the source domain, whereas the user rating vectors $\mat{u}_i^s \in \R^{1 \times m}$ describes the rating relationship between the user $i$ and all the items of the source domain. Accordingly, the item and user rating vectors $\mat{m}_i^t$, $\mat{u}_i^t$ represent the corresponding rating relations of the target domain. 

Taking into consideration that both domains share the same items or users, our primary goal is to exploit and extract knowledge from the source domain and transfer it to the target domain; this way it is possible to make recommendations for items (or users) with no ratings or little information, thus tackling the data sparsity and the cold-start problem in  target domain. In more detail, this task can be seen as a domain adaptation procedure (transfer learning) \cite{da}, which aims to describe the unknown mathematical relationships between the source and target domains. Nonetheless, by tackling this kind of problem two major questions emerged and need to be answered: (i) what to transfer - which information is beneficial to transfer across the domains; and (ii) how to transfer - which learning procedure could be employed to transfer the knowledge. To this end, we address these crucial questions by developing two novel CDR frameworks based on a coupled autoencoder approach.  
\begin{figure}
\centering
 \includegraphics[width=1\linewidth]{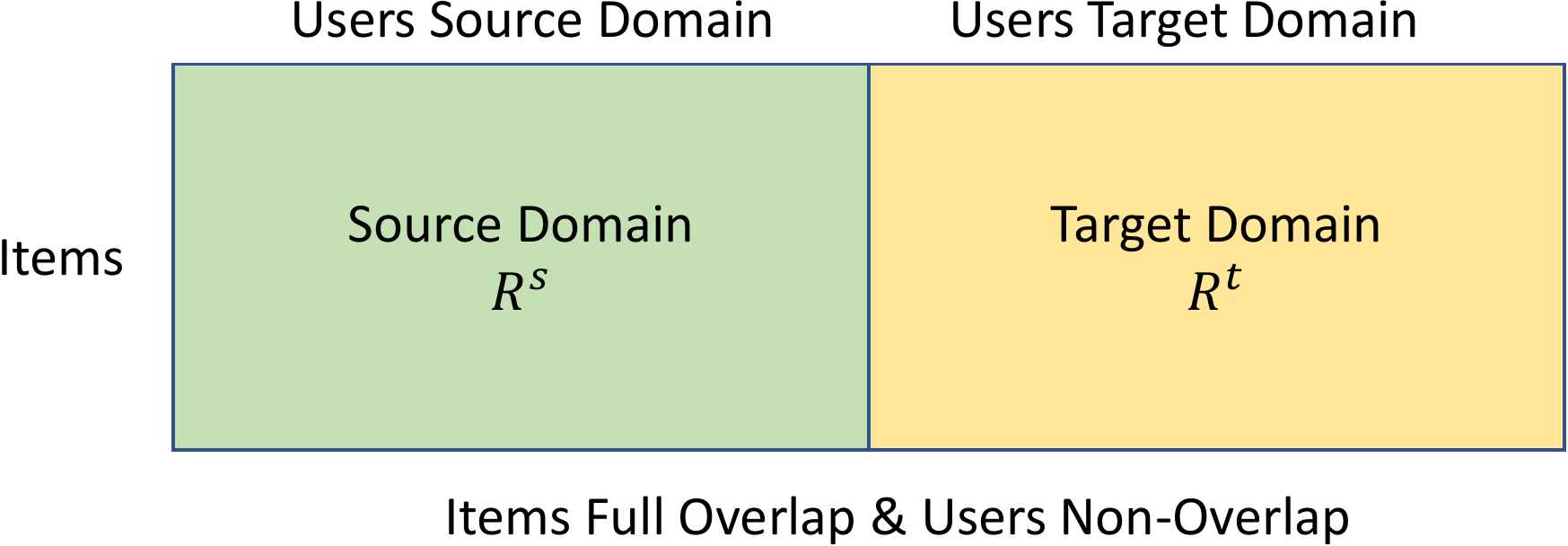}
  \caption{The item-level relevance recommendation task. We assume two related domains, which contain the same items (item full overlap) corresponding to different users (user non-overlap).}
  \label{fig:items_scenario}
\end{figure} 

\begin{figure}
\centering
 \includegraphics[width=1\linewidth]{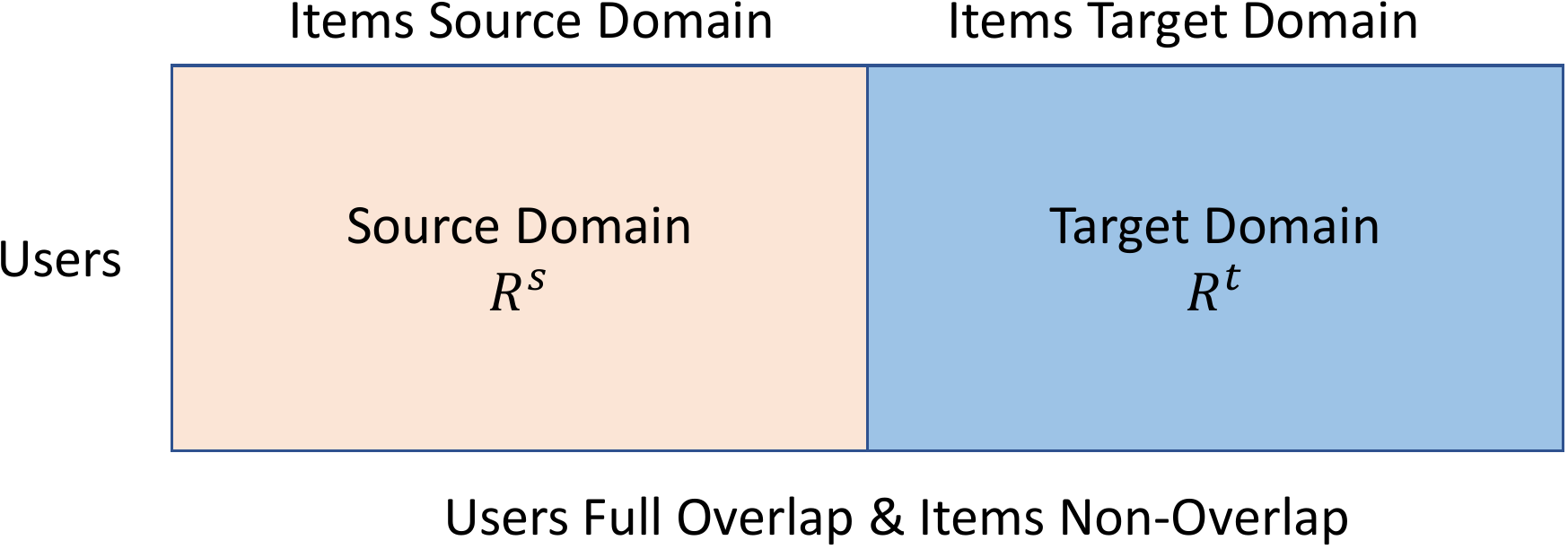}
  \caption{The user-level relevance recommendation task. We assume two related domains, which contain the same users (user full overlap) corresponding to different items (item non-overlap).}
  \label{fig:users_scenario}
\end{figure}

\begin{table}
  \caption{Mathematical Notations}
  \label{tab:commands}
  \begin{tabular}{l l }
    \toprule
    Symbol & Description \\
    \midrule
               & \textit{Source Domain}\\
    
    $n$         & The number of users of the source and target domain \\
    $m$         & The number of items of the source and target domain\\
    $k$ &The dimension of the autoencoder  \\
     &   hidden intrinsic representation   \\
    $\mat{R}^s \in \R^{m \times n}$ & The rating matrix of the source domain\\
    $\hat{\mat{R}^s} \in \R^{m \times n}$ & The predicted rating matrix of the source domain\\
    $\mat{U}^s \in \R^{n \times m}$ & The user rating matrix of the source domain\\
    $\mat{M}^s \in \R^{m \times n}$ & The item rating matrix of the source domain\\
    $\mat{Y}^s_e \in \R^{n \times k}$ & The output of the encoder of $\mat{U}^s$\\
    $\mat{X}^s_e \in \R^{m \times k}$ & The output of the encoder of $\mat{M}^s$\\
    $\hat{\mat{U}^s} \in \R^{n \times m}$ & The output of the decoder of $\mat{Y_e}^s$\\
    $\hat{\mat{M}^s} \in \R^{m \times n}$ & The output of the decoder of $\mat{X_e}^s$\\
    
    \midrule
    & \textit{Target Domain}\\
    $\mat{R}^t \in \R^{m \times n}$ & The rating matrix of the target domain\\
    $\hat{\mat{R}^t} \in \R^{m \times n}$ & The predicted rating matrix of the target domain\\
    $\mat{U}^t \in \R^{n \times m}$ & The user rating matrix of the target domain\\
    $\mat{M}^t \in \R^{m \times n}$ & The item rating matrix of the target domain\\
    $\mat{Y}^t_e \in \R^{n \times k}$ & The output of the encoder of $\mat{U}^t$\\
    $\mat{X}^t_e \in \R^{m \times k}$ & The output of the encoder of $\mat{M}^t$\\
    $\hat{\mat{U}^t} \in \R^{n \times m}$ & The output of the decoder of $\mat{Y_e}^t$\\
    $\hat{\mat{M}^t} \in \R^{m \times n}$ & The output of the decoder of $\mat{X_e}^t$\\

    \bottomrule
    
  \end{tabular}
  \label{tab:notation}
\end{table}

\section{Proposed Models}
\label{sec:models}

In this section, we derive two coupled autoencoder frameworks that can be used for the CDR problem. Specifically, the first one, named CACDR employs a coupled autoencoder method to capture and model the complex relationships between the users and items from the source and target domain, while the second one, named LFACDR can be considered as an extension of the former one utilizing the autoencoders in order to learn in a deep and non-linear manner the user and item-latent factors models in the respective domains. After an initial modeling of domain-specific information in the source, both methods transfer that information to the target domain via a multi-layer perceptron network. 



\subsection{The CACDR Method}

\begin{figure*}
\centering
  \subfloat[Initialization- Item-level relevance scenario.\label{fig:ainit}]{
    \includegraphics[scale=0.52]{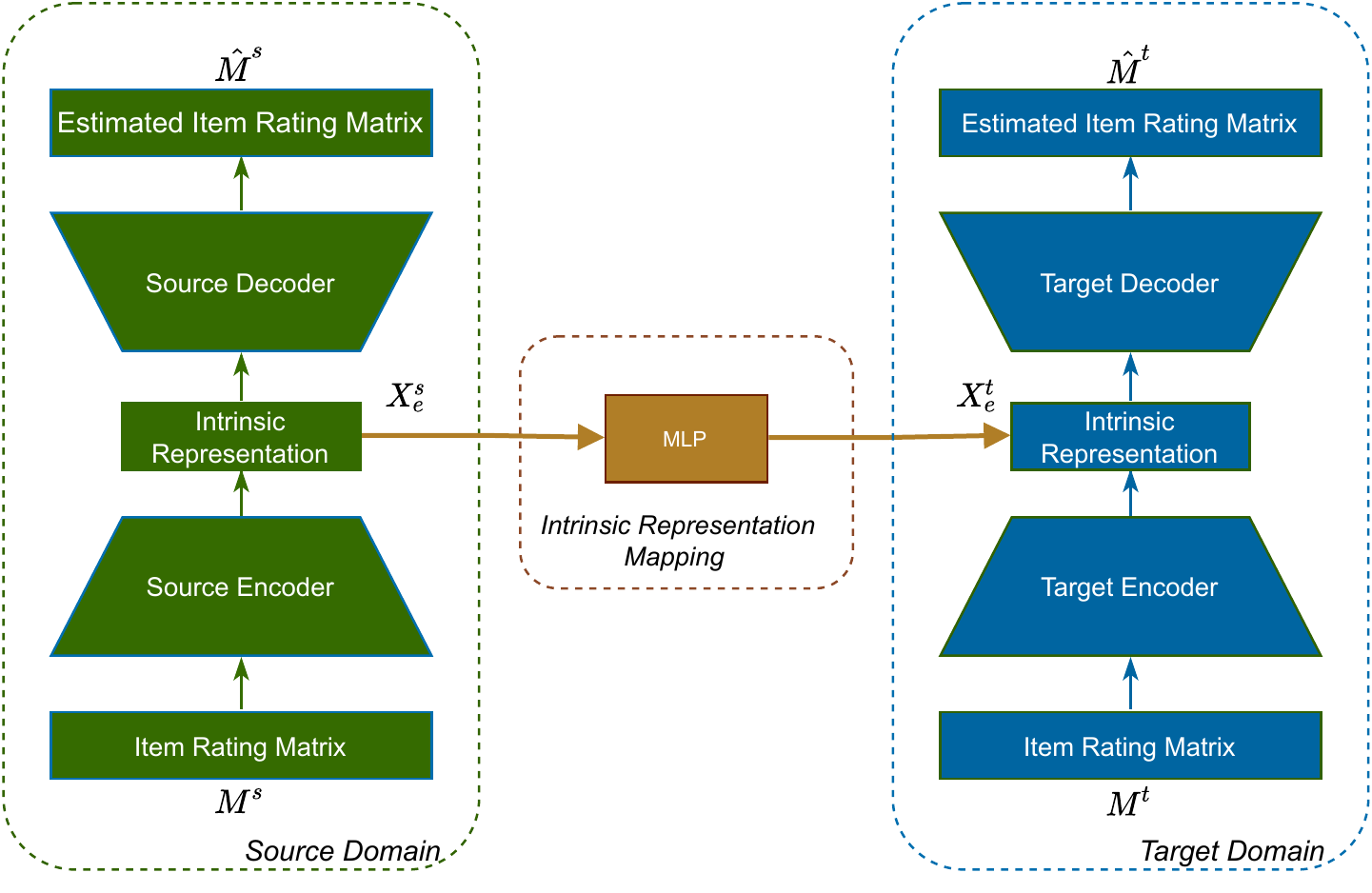}}
    
  \subfloat[Coupled Learning - Item-level relevance scenario.\label{fig:acoupled}]{
    \includegraphics[scale=0.52]{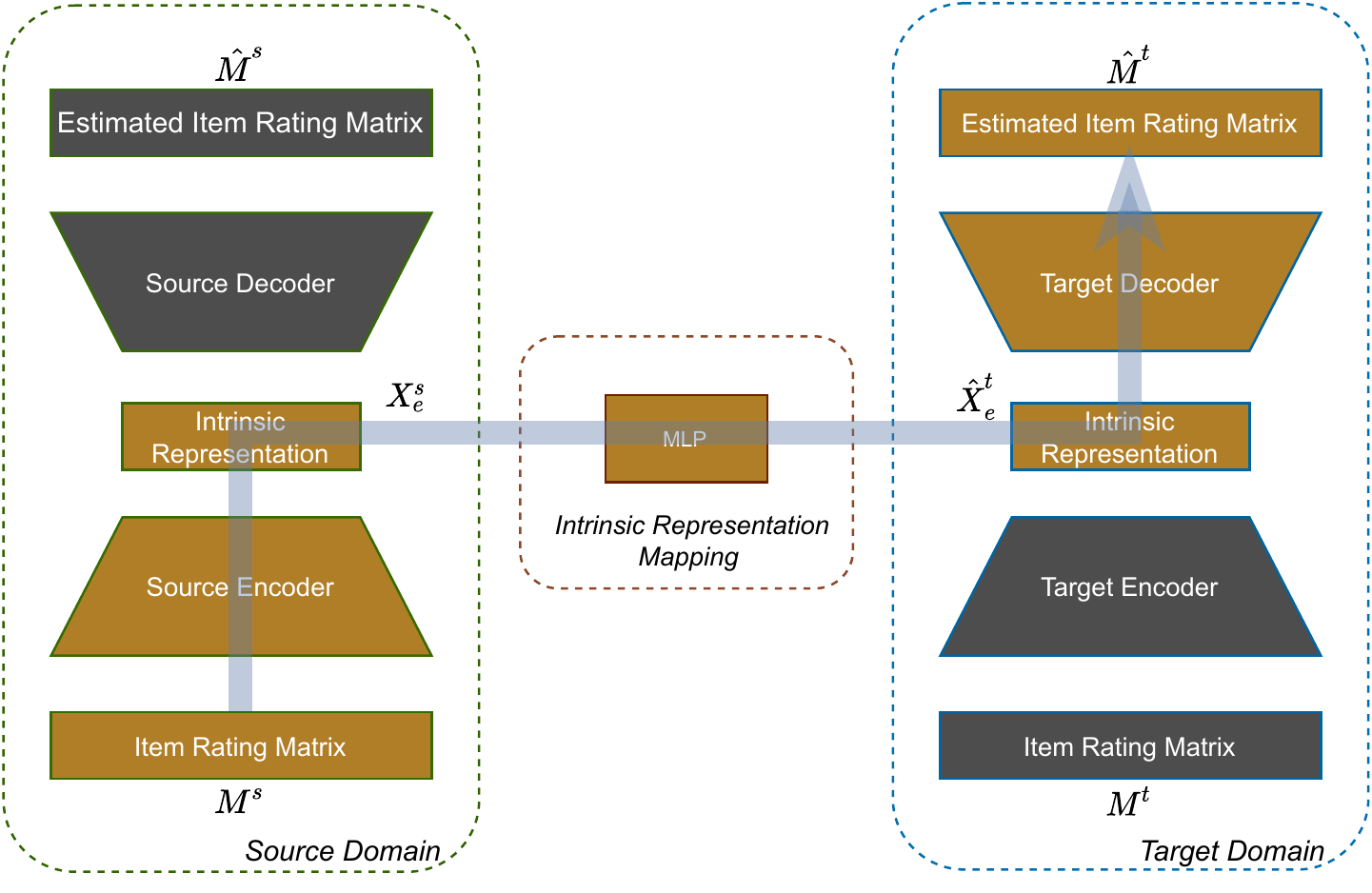}}
  \caption{
  {An illustration of our proposed CACDR model for cross-domain recommendation.(a) Initialization: First the autoencoders are trained to learn the intrinsic representations of the source and target domain (stage 1) and then a mapping function (MLP) is learnt between these representations (stage 2). (b) Coupled Learning: since the autoencoders are trained independently and there is no transfer learning across domains, a coupled autoencoder is employed in order to jointly optimize all the active parts of the autoencoders i.e., the source encoder, the MLP network and the target decoder) involved in the rating prediction in target domain (stage 3). \textbf{Note that we follow similar procedure for the user-level relevance task using the corresponding User rating matrices in source and target domain}}}
  \label{fig:modelA}
\end{figure*}

Autoencoders have demonstrated ground-breaking performance in the unsupervised feature learning domain. Formally, the autoencoder aims to reveal and describe the intrinsic hidden representation of the input by copying its input to its output \cite{Ca2010}. However, the autoencoder as a single domain procedure produces intrinsic representations based only on the input data, thus ignoring the valuable underlying relationships that exist across multiple domains. On the other hand, the coupled autoencoder model is able to capture these internal relationships and derive better representations as the domains influence each other. In particular, the proposed coupled autoencoder based method for CDR, called CACDR consists of three stages. {The first stage employs two autoencoders that learn separately the intrinsic hidden representations of the item rating matrices (item-level relevance scenario) or user rating matrices (user-level relevance scenario) of the source and target domain, respectively.} The second stage  uses a multi-layer Perceptron network (MLP) to model the relationship across domains by learning a mapping function between the intrinsic representations of the source and target domain.
In the previous two stages the autoencoders and the mapping function are trained independently. So we introduce coupling at the third stage in order to  to capture the underlying complex relationships across domains and transfer beneficial knowledge from one domain to the other during the training procedure. Finally, {recommendations can be made for a new item or user in the target domain based on the intrinsic representations of the same item or user in the source domain. The complete proposed methodology is depicted in Fig. \ref{fig:modelA}}.

\subsubsection{Coupled Autoencoder-Based }
{Focusing on the item-level relevance scenario, we consider  the item rating matrices of the source and target domains denoted as $\mat{M}^s$, $\mat{M}^t$ respectively. Note that for the user-level relevance scenario, we employ the user rating matrices in the source and target domain denoted as $\mat{U}^s$, $\mat{U}^t$. Next, we describe the method only for the item-level relevance task, since the methodology is identical for the user-level relevance scenario as well.  }The corresponding source and target autoencoders, which learn the hidden intrinsic representations of the two  matrices, 
{can be obtained by minimizing the following loss functions:
\begin{align}
    &\mathcal{L}^s_{autoenc} = \norm{\mat{M}^s - \hat{\mat{M}^s}}_F^2 = \norm{\mat{M}^s - \mathcal{D}^s(\mathcal{E}^s(\mat{M}^s))}_F^2 \nonumber \\ 
    &\mathcal{L}^t_{autoenc} = \norm{\mat{M}^t - \hat{\mat{M}^t}}_F^2 = \norm{\mat{M}^t - \mathcal{D}^t(\mathcal{E}^t(\mat{M}^t))}_F^2,
    \label{eq:autoenc}
\end{align}}where $\hat{\mat{M}^s}$, $\hat{\mat{M}^t}$ denote the estimated item rating matrices of the source and the target domain, respectively. Formally, the autoencoder comprises of the encoding $\mathcal{E}(.)$ and decoding $\mathcal{D}(.)$ process.

\textbf{Encoding process}: The intrinsic representation of the source item rating matrix $\mat{M}^s \in \R^{m \times n}$ is given by
\begin{align}
    &\mat{X}^s_e\,\,\,=\mathcal{E}^s(\mat{M}^s) \nonumber\\
    \shortintertext{or equivalently,}\nonumber
    &\mat{X}^s_{e,1} =\varphi(\mat{W}^s_{e,1}\mat{M}^s+{b^s_{e,1}}) \nonumber\\
    &\mat{X}^s_{e,2} =\varphi(\mat{W}^s_{e,2}\mat{X}^s_{e,1}+{b^s_{e,2}}) \label{eq:encoding_s} \\
    &\quad \quad ...\nonumber\\
    &\mat{X}^s_{e}  \,\,\, =\varphi(\mat{W}^s_{e,L}\mat{X}^s_{e,L-1}+{b^s_{e,L}}), \nonumber
\end{align}
where $\mat{W}^s_{e,i}$, $b^s_{e,i}$ ($i=1,\ldots,L$) denote the weight matrices and the bias terms for the encoding layers of the source autoencoder, $\varphi(.)$ is the activation function ReLU, $L$ stands for the number of hidden layers, $\mat{X}^s_e\in \R^{m \times k}$ is the output of the source encoder $\mathcal{E}{^s}(.)$ and $k\ll n$.

Similarly, the intrinsic representation of the target item rating matrix $\mat{M}^t \in \R^{m \times n}$ can be defined as
\begin{align}
    &\mat{X}^t_e\,\,\,=\mathcal{E}^t(\mat{M}^t) \nonumber\\
    \shortintertext{or equivalently,}
    &\mat{X}^t_{e,1} =\varphi(\mat{W}^t_{e,1}\mat{M}^t+{b^t_{e,1}}) \nonumber\\
    &\mat{X}^t_{e,2} =\varphi(\mat{W}^t_{e,2}\mat{X}^t_{e,1}+{b^t_{e,2}}) \label{eq:encoding_t} \\
    &\quad \quad ...\nonumber\\
    &\mat{X}^t_{e}  \,\,\, =\varphi(\mat{W}^t_{e,L}\mat{X}^t_{e,L-1}+{b^t_{e,L}}), \nonumber
\end{align}
where $\mat{W}^t_{e,i}$, $b^t_{e,i}$ ($i=1,\ldots,L$) denote the weight matrices and the bias terms for the encoding layers of the target autoencoder, $\varphi(.)$ is the activation function ReLU, $L$ stands for the number of hidden layers, $\mat{X}^t_e\in \R^{m \times k}$ denotes the output of the target encoder $\mathcal{E}{^t}(.)$ and $k\ll m$.


\textbf{Decoding process}:
Accordingly, the reconstructed item rating matrices $\hat{\mat{M}^s}$, $\hat{\mat{M}^t}$ can be derived by
\begin{equation}
   \hat{\mat{M}}^s=\mathcal{D}^s(\mat{X}^s_e)
    \label{eq:decoding_s}
\end{equation}
\begin{equation}
   \hat{\mat{M}}^t=\mathcal{D}^t(\mat{X}^t_e),
    \label{eq:decoding_t}
\end{equation}
where $\mathcal{D}^s(.)$ denotes the decoding procedure of the source autoencoder and $\mathcal{D}^t(.)$ is the decoding procedure of the target autoencoder. Note that both decoders consist also of $L$ fully connected layers. 


To put in a nutshell, the encoding procedure aims to learn a concrete representation of the input in order to capture the complex relationships between the items and users. On the other hand, the decoding process seeks to decode the hidden representations back to the original rating matrices. The accurate decoding procedure enables the autoencoder to learn the rating patterns between items and users and make rating predictions for new items and users.
\subsubsection{Non Linear Mapping}

After obtaining the intrinsic representations of the item (or user) rating matrices, an MLP is employed to capture and model the underlying relationship between the intrinsic representations of the source and target domain ($\mat{X}^s_e$, $\mat{X}^t_e$), thus transferring the appropriate knowledge from the source to target domain. Mathematically, the non linear mapping function $\mathcal{F}(.)$ can be written as

\begin{align}
    &\hat{\mat{X}^t_e}\,\,\,=\mathcal{F}(\mat{X}^s_e) \nonumber \\
    \shortintertext{or equivalently,}
    &\mat{X}^t_{e,1} =\varphi(\mat{W}_{p,1}\mat{X}^s_{e}+{b_{p,1}})\nonumber \\
    &\mat{X}^t_{e,2} =\varphi(\mat{W}_{p,2}\mat{X}^t_{e,1}+{b_{p,2}}) \label{eq:mlp}\\
    &\quad \quad ...\nonumber\\
    &\hat{\mat{X}^t_{e}}  \,\,\, =\varphi(\mat{W}_{p,L}\mat{X}^t_{e,L-1}+{b_{p,L}}) \nonumber 
\end{align}
where $\mat{W}_{p,i}$, $b_{p,i}$ ($i=1,\ldots,L$) denote the weight matrices and the bias terms, $\varphi(.)$ is the activation function ReLU, $L$ stands for the number of hidden layers and $\hat{\mat{X}^t_e}$ is the estimated intrinsic representation of the target domain. 
{Thus, the parameters of the MLP network can be learned minimizing the following loss function
\begin{equation}
   \mathcal{L}_{mlp} = \norm{\mat{X}_e^t - \hat{\mat{X}_e^t}}_F^2 = \norm{\mat{X}_e^t - \mathcal{F}(\mat{X}^s_e)}_F^2.
\end{equation}}

\subsubsection{Cross-domain Rating Predictions}
{The goal of the proposed framework is to recommend new items in the target domain leveraging upon the knowledge of the same items belonging in the source domain (item-level relevance scenario) or new users in target domain using information of the same users in source domain (user-level relevance scenario). In particular, given an item or user $j$ in the target domain, the following methodology is used to recover its predicted rating}:
\begin{enumerate}[1.]
    \item The same item or user is found in the source domain and its intrinsic representation is obtained by employing the autoencoder of the source domain (encoding procedure) according to equation (\ref{eq:encoding_s}).
    \item The corresponding intrinsic representation of the item/user in the target domain can be estimated via  the intrinsic representation of the item/or user  in the source domain and the MLP network based on (\ref{eq:mlp}).
    \item Finally, the predicted rating of the item/user in the target domain is recovered based on target autoencoder (decoding procedure) according to (\ref{eq:decoding_t}).
\end{enumerate}

\subsubsection{Coupled Learning}
\label{c1}
The most critical part of the proposed architecture is the optimization and coupling of the autoencoders along with the mapping  function. However, by learning the autoencoders first via relation (\ref{eq:autoenc}) and then the mapping function (based on the estimated intrinsic representations) via relation (\ref{eq:mlp}) may lead to poor performance, since the autoencoders and the mapping function are optimized independently. In other words, there is no transfer or coupled learning between the source domain (i.e., source  rating matrix)  and the target domain (i.e., target rating matrix). This procedure is piecemeal and thus sub-optimal, since there is no influence from one to the other during the training process. Nevertheless, this methodology can be used as \textit{initialization} process of the model. 

In light of the fact that the ultimate goal of the CACDR method is to efficiently predict the item ratings of the target domain (i.e., the item rating matrix, $\mat{M}^t = \mat{R}^t$ in the item-level relevance scenario or the user matrix, $\mat{U}^t$ in the user-level relevance scenario ), the proposed objective function for optimizing jointly the two autoencoders (source and target) 
{along with the MLP network may be written as:
\begin{align}
     &\norm{\mat{M}^t - \hat{\mat{M}^t}}_F^2 \xRightarrow{(\ref{eq:decoding_t})} \norm{\mat{M}^t - \mathcal{D}^t(\hat{\mat{X_e}^t)}}_F^2 \xRightarrow{(\ref{eq:mlp})} \nonumber \\
     &\norm{\mat{M}^t - \mathcal{D}^t(\mathcal{F}(\mat{X^s_e})}_F^2.
\end{align}
Thus, using equation (\ref{eq:encoding_s}),  the corresponding loss function that we seek to minimize during the coupled learning is given by 
\begin{equation}
    \mathcal{L}_{coupled\_learning} = \norm{\mat{M}^t - \mathcal{D}^t(\mathcal{F}(\mathcal{E}^s(\mat{M^s})}_F^2.
    \label{eq:coupledmodel}
\end{equation}}
Note that now in  relation (\ref{eq:coupledmodel}) the source encoder $\mathcal{E}^s(.)$, the mapping neural network $\mathcal{F}(.)$  and the target decoder $\mathcal{D}^t(.)$ are all explicitly involved in the reconstruction of the desired output $\hat{\mat{M^t}}$. \emph{Hence, in order to couple the two autoencoders with the mapping function a \textbf{coupled deep network} is employed, where its first network component is the source encoder, the second network component is the mapping neural network and its final network component is the target decoder}. Fig. \ref{fig:acoupled} illustrates the proposed coupled architecture. Having obtained, the stacked  network architecture the back-propagation algorithm is used to optimize (\ref{eq:coupledmodel}).  Algorithm 1 summarizes the proposed methodology.  

\begin{algorithm}
\caption{: CACDR learning method}
\begin{algorithmic}[1]
\REQUIRE {The item  rating matrices} of the source and target domain $\mat{M}^s \in \R^{m \times n}$, $\mat{M}^t \in \R^{m \times n} $ 
for the item-level relevance scenario, or the item  user rating matrices $\mat{U}^s \in \R^{n \times m}$, $\mat{U}^t \in \R^{n \times m} $ for the user-level relevance scenario 
\ENSURE \,\,\,The predicted rating matrix of the target domain $\hat{\mat{R}^t}$\\
\COMMENT{\textit{\textbf{Stage A}: Initialization}}
\STATE Initialize the source domain autoencoder by learning the intrinsic representation $\mat{X}^s_e \in \R^{m \times k}$ of the matrix $\mat{M}^s$(item-level relevance scenario) or $\mat{U}^s$ (user-level relevance scenario) via (\ref{eq:encoding_s}).
\STATE Initialize the target domain autoencoder by learning the intrinsic representation $\mat{X}^t_e \in \R^{m \times k}$ of the matrix $\mat{M}^t$(item-level relevance scenario) or $\mat{U}^t$ (user-level relevance scenario) via (\ref{eq:encoding_t}).
\STATE Initialize the MLP network by learning the mapping function from $\mat{X}^s_e$ to $\mat{X}^t_e$ via (\ref{eq:mlp}).\\
\COMMENT{\textit{\textbf{Stage B}: Coupled Learning}}
\STATE Construct the CACDR model via (\ref{eq:coupledmodel}). 
\end{algorithmic}
\end{algorithm}

\subsection{The LFACDR Method}

\begin{figure*}
\centering
  \subfloat[Initialization.\label{fig:binit}]{
    \includegraphics[scale=0.5]{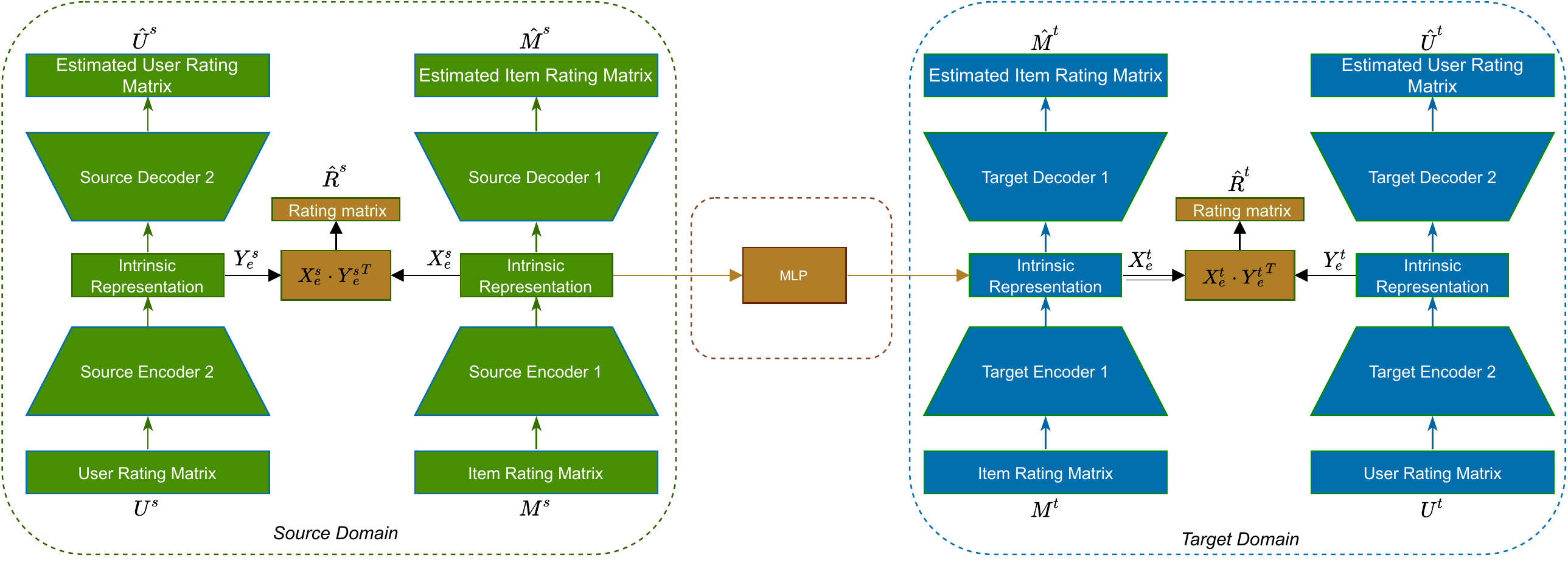}}
    
  \subfloat[Coupled Learning - Item-level relevance scenario.\label{fig:bcoupled}]{
    \includegraphics[scale=0.5]{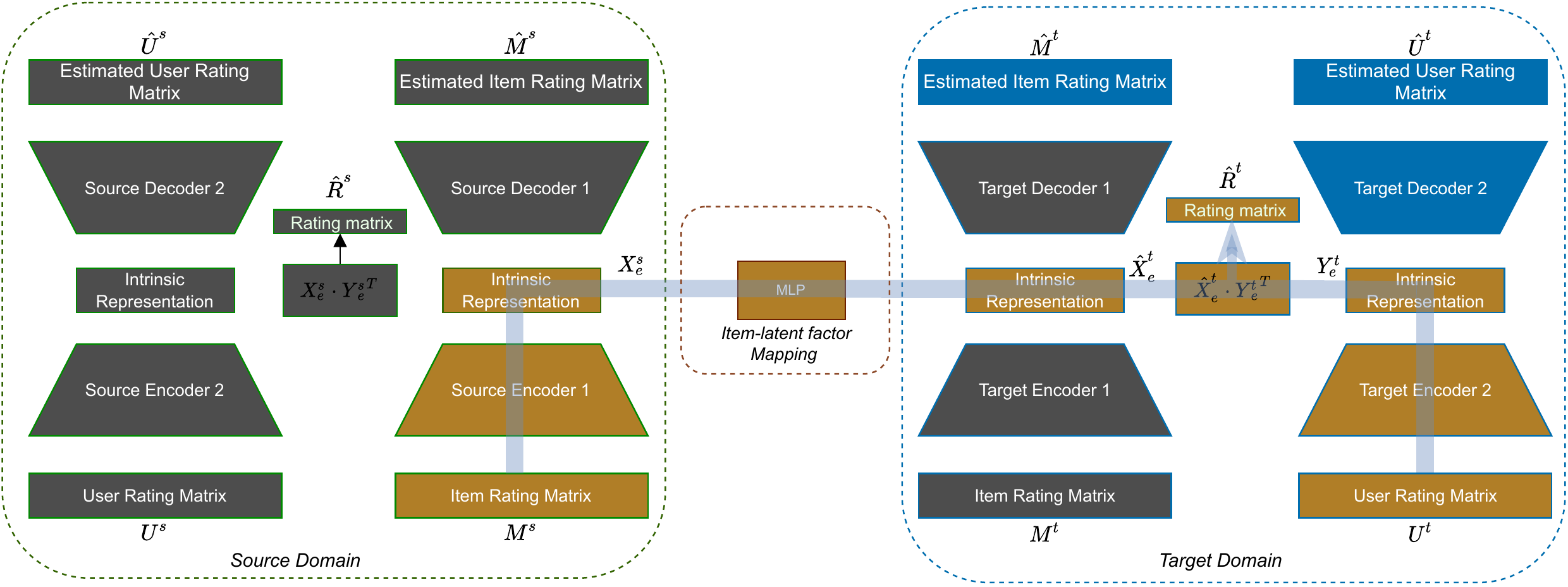}}
    
    \subfloat[Coupled Learning - User-level relevance scenario.\label{fig:bcoupled2}]{
    \includegraphics[scale=0.5]{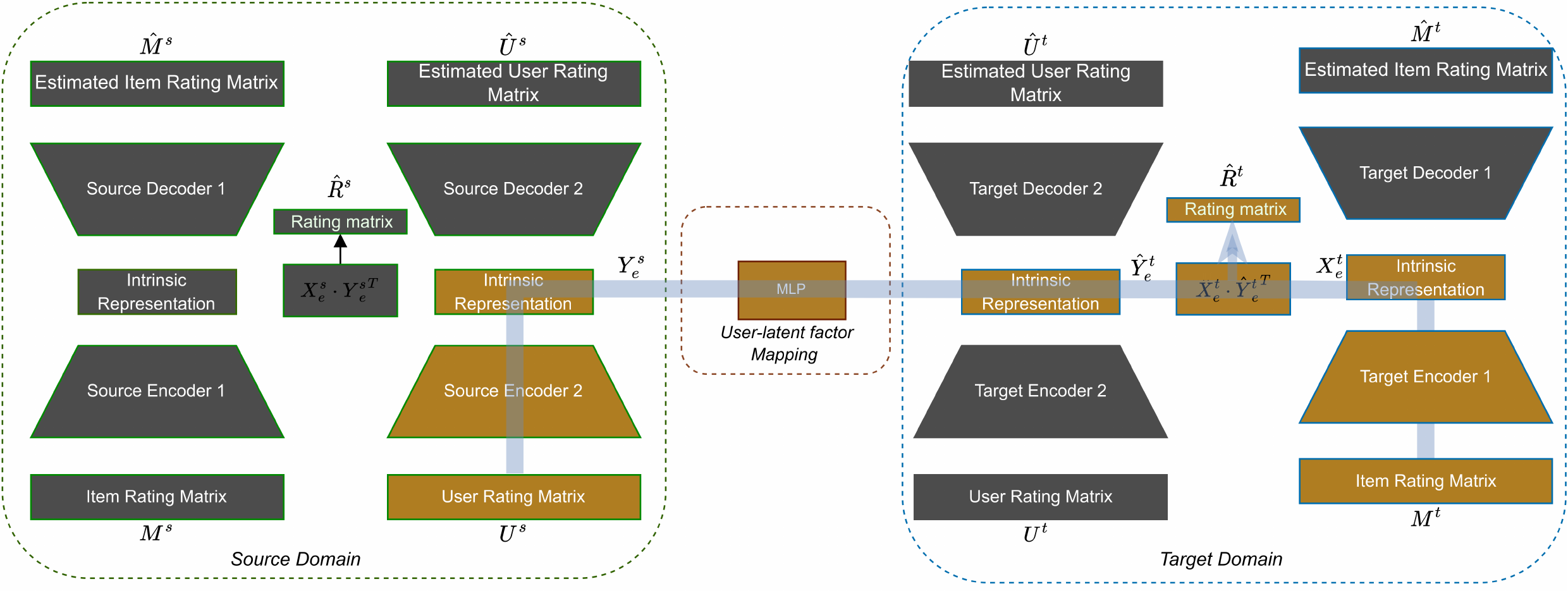}}
  \caption{ An illustration of the proposed LFACDR model for cross-domain recommendation.(a) Initialization: First, the autoencoders are trained to obtain the item and user-latent factors of the source and target domain (stage 1) and then a mapping function (MLP) is learnt between the item latent factor matrices of the source and target domain (stage 2).
  { (b) Coupled Learning: A coupled autoencoder model is employed in order to jointly optimize all the active parts of the autoencoders (i.e., the Source Encoder 1, the MLP network and the Target Encoder 2) involved in the rating prediction in target domain (stage 3).}
  (c) Similar to (b) for the user-level relevance scenario.}
  \label{fig:modelb}
\end{figure*}

As previously mentioned, the autoencoders constitute an ideal mathematical tool to reveal and learn complex low dimensional representations while at the same time they preserve the underlying structure of the input data. This consideration motivates the ensuing cross  domain recommendation methodology that a joint optimization problem is proposed in order to recover in a deep and non-linear manner the user and item-latent factors in both source and target domains. To this end, for each domain two autoencoders are employed to jointly learn the intrinsic representations of user and item rating matrices and decompose the rating matrix into two low-rank matrices, that is the user and item-latent factor matrix. The proposed framework is shown in Fig. \ref{fig:modelb}. 

\subsubsection{Latent Factor Modeling based on Autoencoders}
Consider the user rating matrix $\mat{U}^s \in \R^{n \times m}$, the item rating matrix $\mat{M}^s \in \R^{m \times n}$ and the rating matrix $\mat{R}^s \in \R^{n \times m}$ of the source domain. To obtain the user and item-latent factors of the source domain the following joint constrained optimization problem is proposed, which includes one autoencoder for the items and one autoencoder for the users. 
{Hence, the proposed optimization problem is formulated as   
\begin{align}
  \label{eq:latents}
  \mathcal{L}^s=
  &\norm{\mat{M}^s - \mathcal{D}{^s_m}(\mat{X^s_e})}_F^2 + \norm{\mat{U}^s - \mathcal{D}{^s_u}(\mat{Y^s_e})}_F^2 + \nonumber\\
  &\norm{\mat{Y^s_e}- \mathcal{E}{^s_u}(\mat{U^s})}_F^2 \, \, \,\,+
  \norm{\mat{X^s_e} - \mathcal{E}{^s_m}(\mat{M^s})}_F^2+\\
  & \lambda \norm{\mat{X^s_e}\,\mat{Y^s_e}^T - \mat{R^s}}_F^2,
\nonumber
\end{align}}where $\mathcal{D}{^s_m}(.)$, $\mathcal{E}{^s_m}(.)$ denote the decoding and encoding procedure of the first source autoencoder (items),
$\mat{X^e_s} \in \R^{m \times k}$ is the item-latent factor matrix  derived from the output of the first encoder,   $\mathcal{D}{^s_u}(.)$,$\,$ $\mathcal{E}{^s_u}(.)$ stand for the decoding and encoding procedure of the second source autoencoder (users) and $\mat{Y^e_s} \in \R^{n \times k}$ denotes the user-latent factor matrix derived from the output of the second encoder. Note that the encoding and decoding processes of two autoencoders can be written alternatively according to equations (\ref{eq:encoding_s}) and (\ref{eq:decoding_s}). 

The two autoencoders aim to learn jointly the intrinsic representations of the item and user rating matrices. At the same time, through the proposed constraint optimization problem these intrinsic representations that is the $\mat{X}^s_e$ and  $\mat{Y}^s_e$ act as the desired item and user-latent factor matrices, respectively. Note that similar procedure can be employed to derive the corresponding item and user-latent factor matrices $\mat{X}^t_e$ and  $\mat{Y}^t_e$ of the target domain.

\subsubsection{Non Linear Mapping}
Having acquired the latent factor matrices \{${\mat{X}^e_s, \mat{Y}^e_s,\mat{X}^e_t, \mat{Y}^e_t}$\}  of items and users in the source and target domain, similar to the previous proposed method,  {an L-layer MLP is again used to learn the  mapping function  between the item latent factor matrices of the source and target domain ($\mat{X}^s_e$, $\mat{X}^t_e$) for the item-level relevance scenario or the   the user latent factor matrices of the source and target domain ($\mat{Y}^s_e$, $\mat{Y}^t_e$) for the user-level relevance scenario}. 
Concerning, the item-level relevance scenario, this can be expressed as follows,
\begin{equation}
    \hat{\mat{X}^t_e}\,\,\,=\mathcal{F}(\mat{X}^s_e) \label{eq:mlp_two}
\end{equation}
Similarly,  the parameters of the MLP network can be learned solving the following loss function
\begin{equation}
   \mathcal{L}_{mlp} = \norm{\mat{X}_e^t - \hat{\mat{X}_e^t}}_F^2 = \norm{\mat{X}_e^t - \mathcal{F}(\mat{X}^s_e)}_F^2.
\end{equation}

 Note that for the user-level relevance scenario, the mapping function is defined as follows  
 
 \begin{equation}
    \hat{\mat{Y}^t_e}\,\,\,=\mathcal{F}(\mat{Y}^s_e) \label{eq:mlp_two_user}
\end{equation}
\subsubsection{Rating prediction}
{In general, given a new item (item-level relevance task) or a new user (user-level relevance task) in the target domain with little information, we are not able to calculate an accurate latent factor for making recommendation. In light of this, the corresponding latent factor is learnt from the source domain and a latent factor is derived for the same item or same user in the target domain via the mapping function (\ref{eq:mlp_two})}. 
Concerning, the item-level relevance task the predicted rating between item $i$ and user $j$ in the target domain is given by the following relation,
\begin{equation}
    \hat{\mat{R}^t}(i,j)=\hat{\mat{X}^t_e}(i,:)\, \mat{Y}^t_e(j,:)^T
\end{equation}
where $\hat{\mat{X}^t_e}(i,:) \in \R^{1 \times k}$ denote the estimated item-latent factor of  item $i$ in the target domain based on the corresponding item-latent factor of item $i$ in the source domain via relation (\ref{eq:mlp_two}) and ${\mat{Y}^t_e}(i,:) \in \R^{1 \times k}$  stands of the row vector of matrix ${\mat{Y}^t_e}$, representing the user-latent factor of user $j$ in the target domain.

{Similarly, regarding the user-level relevance the predicted rating between item $i$ and user $j$ in the target domain is given by the following relation,
\begin{equation}
    \hat{\mat{R}^t}(i,j)={\mat{X}^t_e}(i,:)\, \hat{\mat{Y}}^t_e(j,:)^T
\end{equation}}
where $\hat{\mat{Y}^t_e}(j,:) \in \R^{1 \times k}$ denote the estimated user-latent factor of  user $j$ in the target domain based on the corresponding user-latent factor of user $j$ in the source domain via the mapping function in relation (\ref{eq:mlp_two})

\subsubsection{Coupled Learning}
\label{c2}
Similar to the previous procedure the training can be divided into two phases:  the first phase is the initialization and the second phase is the coupled learning.
\textit{Initialization}: First, the autoencoders are trained to obtain the item and user-latent factors of the source and target domain and then  the mapping function is learnt.\\
{
\textit{Coupled learning - Item-level relevance scenario}:
Since the aim of the proposed method in this scenario is to accurately predict the ratings of new items  in the target domain, the objective function for jointly optimizing the autoencoders extracting the item and user-latent factor matrices  
along with the MLP network is given by 
\begin{align}
    \norm{\mat{R}^t - \hat{\mat{R}^t}}_F^2 \Rightarrow &\norm{\mat{R}^t - \hat{\mat{X}^t_e}\, (\mat{Y}^t_e)^T}_F^2 \xRightarrow{(\ref{eq:mlp_two})} \nonumber\\ &\norm{\mat{R}^t - \mathcal{F}(\mat{X}^s_e)\, (\mat{Y}^t_e)^T}_F^2
\end{align}
where 
    $\mat{X}^s_e= \mathcal{E}{^s_m}(\mat{M}^s)$ and $\mat{Y}^t_e= \mathcal{E}{^t_u}(\mat{U^t}) \nonumber$ .}
{
Hence, we aim to minimize the following loss function during the coupled learning stage 
\begin{align}
    \mathcal{L}_{coupled\_learning} = \norm{\mat{R}^t - \mathcal{F}(\mat{X}^s_e)\, (\mat{Y}^t_e)^T}_F^2 +\nonumber\\ \norm{\mat{X}^s_e- \mathcal{E}{^s_m}(\mat{M}^s)}_F^2 + \norm{ \mat{Y}^t_e- \mathcal{E}{^t_u}(\mat{U^t})}_F^2.
    \label{eq:obj2}
\end{align}
From relation (\ref{eq:obj2}) it is easy to verify that the source encoder for the items $\mathcal{E}{^s_m}(.)$, the target encoder for the users $\mathcal{E}{^t_u}(.)$ and MLP network are all explicitly involved in the reconstruction of the desired output $\hat{\mat{R}^t}$. Thus, the three network units can be coupled together by jointly optimizing them through the back propagation algorithm. }
{
\textit{Coupled learning - User-level relevance scenario}:
The
aim of the proposed method in this scenario is to accurately
predict the ratings of new users in the target domain, the
objective function for jointly optimizing the autoencoders
extracting the item and user-latent factor matrices along with
the MLP network is given by
\begin{align}
    \norm{\mat{R}^t - \hat{\mat{R}^t}}_F^2 \Rightarrow &\norm{\mat{R}^t - {\mat{X}^t_e}\, (\hat{\mat{Y}}^t_e)^T}_F^2 \xRightarrow{(\ref{eq:mlp_two})} \nonumber\\ &\norm{\mat{R}^t - \mat{X}^t_e\, (\mathcal{F}(\mat{Y}^s_e))^T}_F^2
\end{align}
where 
    $\mat{X}^t_e= \mathcal{E}{^t_m}(\mat{M}^t)$ and $\mat{Y}^s_e= \mathcal{E}{^s_u}(\mat{U^s}) \nonumber$ .
Thus, the loss function during this stage is 
\begin{align}
    \mathcal{L}_{coupled\_learning} = \norm{\mat{R}^t - \mat{X}^t_e\, (\mathcal{F}(\mat{Y}^s_e))^T}_F^2 +\nonumber\\ \norm{\mat{X}^t_e- \mathcal{E}{^t_m}(\mat{M}^t)}_F^2 + \norm{ \mat{Y}^s_e- \mathcal{E}{^s_u}(\mat{U^s})}_F^2.
    \label{eq:obj2_u}
\end{align}}

Fig. \ref{fig:bcoupled} demonstrates the proposed coupled framework.
The overall methodology is summarized in Algorithm 2.

\begin{algorithm}
\caption{: LFACDR learning method}
\begin{algorithmic}[1]
\REQUIRE The item rating matrices of the source and target domain $\mat{M}^s \in \R^{m \times n}$, $\mat{M}^t  \in \R^{m \times n}$ and corresponding user rating matrices $\mat{U}^s \in \R^{n \times m}$, $\mat{U}^t  \in \R^{n \times m}$
\ENSURE \,\,\,The predicted rating matrix of the target domain $\hat{\mat{R}^t}$\\
\COMMENT{\textit{\textbf{Stage A}: Initialization}}
\STATE Initialize the two source domain autoencoders by learning the item-latent factor matrix $\mat{X}^s_e \in \R^{m \times k}$ and the user-latent factor matrix $\mat{Y}^s_e \in \R^{n \times k}$ via (\ref{eq:latents}).

\STATE Initialize the two target domain autoencoders by learning the item-latent factor matrix $\mat{X}^t_e \in \R^{m \times k}$ and the user-latent factor matrix $\mat{Y}^t_e \in \R^{n \times k}$ via (\ref{eq:latents}).

\STATE Initialize the MLP network by learning the mapping function from $\mat{X}^s_e$ to $\mat{X}^t_e$ for the the item-level relevance scenario via (\ref{eq:mlp_two}) or  $\mat{Y}^s_e$ to $\mat{Y}^t_e$ for the user-level relevance scenario via (\ref{eq:mlp_two_user}) .\\
\COMMENT{\textit{\textbf{Stage B}: Coupled Learning}}

\STATE Construct the LFACDR model via (\ref{eq:obj2}) for the item-level relevance scenario or (\ref{eq:obj2_u}) for the user-level relevance scenario. 
\end{algorithmic}
\end{algorithm}

\section{Experimental validation}
\label{sec:exp}

To validate the efficacy and applicability of the two proposed methods, extensive experiments were conducted in the context of the CDR problem. In particular, this study examines two CDR scenarios: the first one  focuses on an item-level relevance cross-domain recommendation scenario, where two domains (source and target) share common items (e.g., movies) and  contain different users; the second scenario examines a user-level relevance cross-domain recommendation scenario, where the source and target domain share common users and contain different items e.g., (the source domain may contain movies as items and the target domain may contain books as items) .  
{Our goals are to demonstrate:  
\begin{itemize}
    \item that the proposed methods are capable to exploit and transfer valuable information from the source domain to improve the recommendation performance in the target domain, thus tackling effectively the cold start problem, in which items (e.g., movies) or users in target domain have no historic information. 
    \item the effect of  different sparsity levels of the rating matrices in the source and target domains.
    \item the performance of the methods in comparison to other state-of-the-art approaches.
\end{itemize}}

\subsection{Datasets}

\textbf{Item-level relevance CDR scenario - Datasets}: Two publicly available benchmark datasets were employed to demonstrate the merits of the proposed recommendation frameworks. Namely, we used the MovieLens\footnote{https://grouplens.org/datasets/movielens/} and Netflix\footnote{https://www.kaggle.com/netflix-inc/netflix-prize-data} datasets, which contain a large portion of same movies, thereby forming an item-level relevance scenario, as depicted in Figure \ref{fig:items_scenario}. Note that  users in the two platforms are different. According to IMDB information more than $6000$ movies are the same across MovieLens and Netflix datasets.

\textbf{User-level relevance CDR scenario - Datasets}: Concerning the CDR scenario with the shared users, we employed the Douban dataset \cite{man}, where users provide ratings to different items such as movies, books and music, thus forming a natural user-level relevance scenario, as depicted in Figure \ref{fig:users_scenario}. 

\subsection{Experimental Setup}

\textbf{Experimental Setting}:

 Regarding the \textbf{item-level relevance CDR scenario} for completeness purposes, first we took the MovieLens as source domain and Netflix as target domain and then we used  the Netflix  as source domain and the MovieLens as target domain. 
Taking into account the huge number of users in both domains (i.e., more than $2.000.000$ users in Netflix and more than $150.000$ users in MovieLens platform), we followed the same strategy as in \cite{Gao2019, Zhong2020} and we randomly sub-sampled a certain number of users in the MovieLens and Netflix dataset, ensuring that the selected users had at least 5 interactions with the considered movies in the  domain that belong. In order to examine the effect of different sparsity levels of the rating matrices, we repeated the above procedure  four times, selecting randomly  different  users with varying number of interactions. Hence, we ended up with four data subsets (i.e., No 1, No 2, No.3, No. 4) where the sparsity level increase from data subset No. 1 to data subset No. 4.  Table \ref{tab:data_stats} provides the  detailed statistics of the four resulting data subsets in which the number of the common movies, the number of the selected users and the sparsity of the rating matrices are listed.

Regarding the \textbf{user-level relevance CDR scenario}, first we took the book as source domain and the movie as target domain and then we employed the music as source domain and the book as target domain.  The datasets are shown in Table \ref{tab:data_stats_duban}.



\begin{table}
  \caption{
  {Statistics of the four Data Subsets derived from the Movielens and Netflix datasets.} }
   \centering
  \resizebox{\linewidth}{!}{
  \begin{tabular}{ccccl}
    \toprule
    Data subsets&Domains &\#Common Movies & \#Users& Sparsity \\
    \midrule
    No. 1&MovieLens & 5819 & 40000 & 97.22\% \\
    &Netflix   & 5819 & 40000 & 94.80\% \\
    \midrule
    No. 2&MovieLens & 5986 & 48575 & 98.17\% \\
        &Netflix   & 5986 & 49681 & 97.02\% \\
    \midrule
    No. 3&MovieLens & 6000 & 30100 & 99.35\% \\
    &Netflix   & 6000 & 30040 & 98.99\% \\    
    \midrule
    No. 4&MovieLens & 4200 & 44850 & 99.59\% \\
    &Netflix   & 4200 & 43775 & 99.36\% \\        
    \bottomrule

  \end{tabular}}
  \label{tab:data_stats}
\end{table}

\begin{table}
  \caption{
  {Statistics of the Douban Datasets.} }
   \centering
  \resizebox{\linewidth}{!}{
  \begin{tabular}{ccccl}
    \toprule
    Data subsets&Domains &\#Common Users & \#Items& Sparsity \\
    \midrule
    No. 1&Douban Book & 2120 & 8659 & 99.41\% \\
    &Douban Movie   & 2120 & 16584 & 98.20\% \\
    \midrule
    No. 2&Douban Music & 1580 & 6888 & 99.45\% \\
        &Douban Book   & 1580 & 7545 & 99.50\% \\
   
    \bottomrule

  \end{tabular}}
  \label{tab:data_stats_duban}
\end{table}

\textbf{Training and Testing Setting}: 
\textit{Item-level scenario}: In our extensive experiments the four data subsets presented in Table \ref{tab:data_stats} were randomly divided into the training set ($80\%$) and the testing set ($20\%$). In more detail, the $80\%$ of the common movies was used during the training phase and the rest $20\%$ was used during the testing.  Regarding the testing set, we removed all the rating information from the movies in the target domain, thus considering these entities as \textbf{cold-start} movies. The testing movies from the source domain were employed in order to transfer valuable information to the target domain.  Furthermore, taking into account that different training and testing sets may affect the recommendation performance, this splitting procedure was repeated $10$ times and the average results were reported. Finally, it should be noted that we normalized the scale of ratings between 0 and 1 following the same strategy as in \cite{Li2020}. \textit{User-level scenario}: Similar procedure was employed. $80\%$ of the common users was used during the training phase and the rest $20\%$ was used during testing. 

\textbf{Parameter Settings}: The parameters of our proposed methods are determined to be ideal via exploration of the parameter space. For \textbf{CACDR} the sizes of the autoencoders layers were set to \{256, 128, 64, 64, 128, 256\}, the sizes of the MLP layers are set to \{64, 128\} and the batch size was 32. Regarding the training of the CACDR method for the initialization stage the number of epochs is 250, the learning rate was set to $10^{-3}$ and the  $l$-2 regularizer term was $10^{-5}$. Accordingly, for the coupled learning stage the number of epochs was set to 300, the learning rate and regularization term was $10^{-5}$. For \textbf{LFACDR } the sizes of the autoencoders layers were set to \{512, 256, 128, 128, 256, 512\}, the sizes of the MLP layers were set to \{128, 256\} and the batch size was 500. Furthermore, for the initialization training stage the number of epochs was set to 250, while the learning rate and the regularizer term were $10^{-3}$ and $10^{-5}$, respectively. Additionally, for the coupled training stage the number of epochs was set to 300 and the learning rate was $10^{-5}$, while the regularizer term remained constant. Finally, the Adam optimizer was employed to train the proposed models and the ReLU was used as activation function. 

\textbf{Loss function and Evaluation metrics}:
Concerning the loss function, in this study we employed the Masked Root Mean Squared Error loss following the approach from \cite{Kuchaiev2017}, since the zero values should be ignored during the training stage of the proposed models. In addition, to evaluate the recommendation performance we adopted the Root Mean Squared Error (RMSE) and Mean Absolute Error (MAE) metrics.

\textbf{Compared methods}: To showcase the added value of the proposed methods (CACDR and LFACDR), we compared with the following CDR frameworks:
\begin{itemize}
    \item[$\diamond$]\textbf{DDTCDR}\cite{Li2020}: It is a recent state-of-the-art CDR model. In more details, DDTCDR exploits the merits of the dual transfer learning and the feature embedding method to transfer knowledge across domains. Furthermore it employs an orthogonal mapping to preserve user relations in latent space.
    \item[$\diamond$]\textbf{DARec}\cite{Darec}: This state-of-the-art framework employs a deep domain adaption model to extract and transfer patterns from rating matrices in different domains, without considering any auxiliary information. 
    \item[$\diamond$]\textbf{EMCDR}\cite{man}: This CDR framework utilizes a matrix factorization methodology to learn the latent factors and then a multi-layer perceptron is used to model the mapping function between the latent factors of the source and the target domain. This method provides four frameworks and we chose the two best ones, namely the MF\_EMCDR\_MLP employing MF (matrix factorization) as its latent factor modeling and the BPR\_EMCDR\_MLP employing BPR (bayesian personalized ranking) as its latent factor modeling.
    \item[$\diamond$]\textbf{CST}\cite{Pan2010}: It compacts the sparsity problem and enhances the recommendation performance by transferring the latent factors obtained from the source domain into the target domain. The model employs matrix factorization to deduce the user and item-latent factors in the source domain, and transfer them into the target domain via a regularization method.
    \item[$\diamond$]\textbf{LFM}\cite{lfm}: It uses a collective matrix factorization method exploiting correlated information across domains via Localized factor models. Each user and item has a global latent factor common across domains.
\end{itemize}
\subsection{Performance Evaluation - Item-level and User-level relevance  scenario}

{Table \ref{tab:results_1} and Table \ref{tab:results_netflixTomovielens} summarize the average \textbf{item-level} relevance recommendation performance results in terms of RMSE and MAE for the four examined data subsets presented in Table \ref{tab:data_stats}.} Additionally, Table \ref{tab:results_douban} summarizes the \textbf{user-level} relevance recommendation performance results (i.e., common users across the source and target domain) for the examined datasets given in Table \ref{tab:data_stats_duban} using firstly the Book domain as source domain and the Movie domain as target domain and next using the Music domain as source domain and the Book domain as target domain. 

It is evident that the proposed coupled autoencoder-based frameworks, (i.e., CACDR and LFACDR) gave better results than the other CDR methods for both the item-level and user-level relevance recommendation scenarios. Moreover, it is noteworthy that both proposed methods (and especially the LFACDR method) were able to maintain low RMSE and MAE values for different sparsity values of the datasets compared to the other baseline models, where their performance degraded for high levels of sparsity level.

\textbf{Comparison against shallow learning  approaches.}
{The proposed models significantly outperform the  CST,  LFM and EMCDR approaches. In more detail, the above approaches utilize matrix factorization techniques to obtain the latent factor models, thus these methods can only capture rather shallow and linear characteristics from the datasets compared to our models that employ deep coupled autoencoders allowing them to capture more complex and non-linear features from the collaborative relationships of the users and items. Furthermore, another major difference between the proposed models and the EMCDR methods is that, the EMCDR approaches learn the latent factors separately from the mapping function. However, this procedure is sub-optimal, since there is no influence between the domains during the learning of the latent factors. On the other hand, the proposed models, utilize the coupled autoencoders to extract the latent representations and transfer valuable information across the domains during the training phase.  }

\textbf{Comparison against deep learning approaches.}
{Although the state-of-the-art baselines that is the DDTCDR and DARec methods exhibit good recommendation results, our proposed models  performed even better and that can be attributed to the following reason. The DDTCDR and DaRec models employ non-linear functions to extract the latent factors (autoencoders)  the learning of the latent factors of the source and target domain along with the mapping function are learnt separately, and hence the transferred knowledge between the domains during the learning stage is rather limited. Different from that, due to the coupled learning stage analyzing in Sections \ref{c1} and \ref{c2}, the autoencoders of the source and the target domain with the mapping function were optimized jointly and the non-linear relations across domains could be transferred much more effectively during the training stage. More details regarding of the impact of coupled learning stage on the recommendation performance of our models that justifies their superiority is given in Section \ref{clr}}.

\textbf{Comparison CACDR and LFACDR.}
 {
Focusing on the proposed methods,
the LFACDR method exhibits better results compared
to the CACDR method in most cases. This finding is mostly
attributed to the fact that the LFACDR method exploits not
only the information of the items but also the information
deriving from the users. However,  it should be highlighted that the CACDR approach is less computationally demanding, since it employs only two autoencoders in the source and target domain. On the other hand,  the LFACDR approach requires four autoencoders in total in order to learn the user and item latent factors in the source and target domain. }

\begin{table*}
\small
  \caption{Item-level relevance recommendation performance using the  MovieLens dataset as source domain and the Netflix dataset target domain}
  \resizebox{\linewidth}{!}{
  \begin{tabular}{ccccccccccl}
    \toprule
    Data Subset&Metrics & CST & LFM & MF\_EMCDR & BPR\_EMCDR& DDTCDR & DARec& CACDR & LFACDR \\
    \midrule
    No. 1&RMSE & 0.2399 & 0.2341 & 0.2289 & 0.2214 & 0.2032& 0.1783  & 0.1772 & \textbf{0.1740} \\
        &MAE   & 0.1877& 0.1819 & 0.1757 & 0.1744 & 0.1648& 0.1426  & 0.1385 & \textbf{0.1360} \\

    \midrule
    No. 2&RMSE & 0.2314 & 0.2228 & 0.2079 & 0.2018 & 0.1956 & 0.1868  & 0.1738 & \textbf{0.1711}  \\
         &MAE   & 0.1813 & 0.1761 & 0.1643 & 0.1617 & 0.1557& 0.1496  & 0.1356  & \textbf{0.1331} \\
        
    \midrule
    No. 3&RMSE & 0.2578  &0.2513 & 0.2387 & 0.2324 & 0.2107& 0.2093  & 0.1980 & \textbf{0.1922} \\
        &MAE   & 0.2181 &  0.2123& 0.1868 & 0.1849 & 0.1699 & 0.1678  & 0.1560 & \textbf{0.1511} \\
    \midrule
    No. 4&RMSE & 0.2673  &0.2617 & 0.2494 & 0.2435 & 0.2229& 
0.2221  & 0.2139 & \textbf{0.2047} \\
        &MAE   &0.2302 &  0.2282& 0.2047 & 0.2009 & 0.1817 & 0.1789  & 0.1687 & \textbf{0.1607} \\
    \bottomrule
  \end{tabular}}
    \label{tab:results_1}
\end{table*}

\begin{table*}
  \caption{
  {Item-level relevance recommendation performance using  the Netflix dataset as source domain and MovieLens dataset as target domain}}
  
  \resizebox{\linewidth}{!}{
  \begin{tabular}{ccccccccccl}
    \toprule
    Data Subset&Metrics & CST & LFM & MF\_EMCDR & BPR\_EMCDR& DDTCDR & DARec& CACDR & LFACDR \\
    \midrule
    No. 1&RMSE & 0.2378 & 0.2233 & 0.2206 & 0.2210 & 0.1998& 0.1824  & 0.1692 & \textbf{0.1683} \\
        &MAE   & 0.1868& 0.1807 & 0.1741 & 0.1740 & 0.1589& 0.1471  & 0.1298 & \textbf{0.1289} \\

    \midrule
    No. 2&RMSE & 0.2302 & 0.2208 & 0.2027 & 0.2002 & 0.1943 & 0.1812  & \textbf{0.1674} & {0.1692}  \\
         &MAE   & 0.1806 & 0.1747 & 0.1624 & 0.1604 & 0.1498& 0.1457  & \textbf{0.1282}  & {0.1301} \\
        
    \midrule
    No. 3&RMSE & 0.2541  &0.2504 & 0.2333 & 0.2301 & 0.2088 & 0.2000  & 0.1932 & \textbf{0.1837} \\
        &MAE   & 0.2153 &  0.2114& 0.1851 & 0.1821 & 0.1644 & 0.1601  & 0.1494 & \textbf{0.1418} \\
    \midrule
    No. 4&RMSE & 0.2624  &0.2612 & 0.2487 & 0.2412 & 0.2174&  0.1999 & 0.1985 & \textbf{0.1937} \\
        &MAE   &0.2287 &  0.2274& 0.2048 & 0.1991 & 0.1695 & 0.1602  & 0.1524 & \textbf{0.1480} \\
    \bottomrule
  \end{tabular}}
    \label{tab:results_netflixTomovielens}
\end{table*}

\section{Ablation Study}

\subsection{Impact of Coupled learning stage} \label{clr}
To demonstrate the impact of the coupled learning on the recommendation performance of the proposed methods, we conducted some experiments with and without the coupled learning stage during the training  procedure of our methods. 
 According to Table \ref{tab:coupledres} and Table \ref{tab:coupled_duban} which summarize the results for the item-level (we observed similar results  using the Netflix dataset as source domain and Movielens domain as target domain) and user-level relevance scenario respectively, the coupled learning notably improves the performance of the proposed methods, thus validating our claims that the coupled autoencoders are able to capture not only the existing relationships in each domain separately, but more importantly to model the underlying relationships between the source and target domains.

\begin{table}
\small
  \caption{
  {The performance  of the proposed models with and without Coupled learning stage using the  MovieLens dataset as source domain and the Netflix dataset target domain} }
\centering
  {
  \begin{tabular}{cccccl}
    \toprule
      Data& Metrics  & CACDR  & CACDR& LFACDR& LFACDR  \\
      &   &  without & with &  without &  with \\
    \midrule
    No. 1&RMSE & 0.1827  & 0.1772  & 0.1849  & 0.1740  \\
        &MAE   & 0.1441 & 0.1385  & 0.1477  & 0.1360   \\
    \midrule
    No. 2&RMSE & 0.1815  & 0.1738 & 0.2101 & 0.1711  \\
        &MAE   & 0.1420  & 0.1356 & 0.1642 & 0.1331   \\
    \midrule
    No. 3&RMSE & 0.2183  & 0.1980  & 0.2094 & 0.1922  \\
        &MAE   & 0.1723  & 0.1560 & 0.1651  & 0.1511  \\  
     \midrule
    No. 4&RMSE & 0.2333  & 0.2139 & 0.2288  & 0.2047  \\
        &MAE   & 0.1860  & 0.1687 & 0.1816  & 0.1607   \\
    \bottomrule
  \end{tabular}}
    \label{tab:coupledres}
\end{table}

\subsection{ Impact of Initialization stage}

As mentioned above the proposed methods consist of two main stages, i.e., the Initialization stage and the Coupled learning stage (see Figures 2 and 3). The initialization stage is a critical component affecting the recommendation performance  of the proposed models. To highlight its importance, we conducted experiments with two training schemes. During the first scheme, called \textit{with Initialization}, the two proposed models trained in two stages, the initialization stage where the involved autoencoders along with the mlp network are trained independently and the coupled learning stage where the autoencoders and the MLP network are optimized end-to-end. During the second scheme, called \textit{without Initialization}, we employed the coupled learning stages of the two proposed methods without the autoencoders and mlp networks are initialized properly (i.e., without the Initialization stage). Table \ref{tab:initres} and Table \ref{tab:init_duban} summarizes the results. It is evident that the initialization stage plays an important role, improving the performance of our methods.   

Considering  the vastness of the  parameter space of a deep learning model, an inappropriate initialization procedure of the proposed models may affect their performance. In other words, since the neural networks are non-convex functions containing a large number of local minima, an improper initialization of the parameters of the models  may lead the optimization process to get stuck on local minima with suboptimal performance. On the other hand, the proposed two stage training process guarantees that during the initialization stage the models will be initialized with proper parameters adapted to the statistical distributions of the training data, and during the coupled learning stage, where the models are optimized end-to-end, these parameters will be preserved allowing the proposed methods to further adapt to the structure of the data and hence improve the performance.

\begin{table}
\small
  \caption{
  {The performance  of the proposed models with and without Initialization  stage using the  MovieLens dataset as source domain and the Netflix dataset target domain} }
\centering
  {
  \begin{tabular}{cccccl}
    \toprule
      Data& Metrics  & CACDR& CACDR& LFACDR & LFACDR    \\
      &   &  without & with &  without &  with \\
    \midrule
    No. 1&RMSE & 0.1801  & 0.1772  & 0.1814 & 0.1740  \\
        &MAE   & 0.1406 & 0.1385  & 0.1435  & 0.1360   \\
    \midrule
    No. 2&RMSE & 0.1769 & 0.1738 & 0.1799 & 0.1711  \\
        &MAE   & 0.1394  & 0.1356 & 0.1398 & 0.1331   \\
    \midrule
    No. 3&RMSE & 0.2066  & 0.1980  & 0.2004 & 0.1922  \\
        &MAE   & 0.1597  & 0.1560 & 0.1591  & 0.1511  \\  
     \midrule
    No. 4&RMSE & 0.2205  & 0.2139 & 0.2126 & 0.2047  \\
        &MAE   & 0.1740  & 0.1687 & 0.1692  & 0.1607   \\
    \bottomrule
  \end{tabular}}
    \label{tab:initres}
\end{table}

\subsection{ Impact of Latent Dimension and Complexity Analysis}

The latent dimension constitutes a crucial factor effecting the efficacy of different cross-domain recommendation models, hence in this experiment the impact of latent dimension $k$ on the proposed models is investigated. In more details, fixing the other parameters of our CDR methods, we examined a broad range of latent dimensions $k$, namely {8, 32, 64, 128, 256}. Table \ref{tab:latent_dimension} summarizes the results. The best results for the CACDR and LFACDR occurred when the latent dimension was set to 64 and 128 respectively. It should be mentioned that the LFACDR method exhibited better performance compared to the CACDR method in most cases. This finding is mostly attributed to the fact that the LFACDR method exploits not only the information of the items but also the information deriving from the users. 
Additionally, from table \ref{tab:latent_dimension} we can deduce that the performance of the proposed models was only slightly affected by the change of the latent dimension, thus indicating their robustness. Similar results, we observed  using
the Douban dataset, exploring the user-level relevance scenario.

{Complexity analysis for the proposed methods and the other compared recommendation approaches is summarized in Table \ref{tab:complexity}. To simplify the analysis, only the main steps of each approach are considered, where the $n$ denote the number of users, $m$ represents the number of items, $k$ denotes the latent dimension, $L$ is the number of the layers of the neural network, and $T$ stands for the number of iterations.}   

\begin{table}
  \caption{
  {Complexity analysis of the proposed methods and the compared baselines .} }
   \centering
  \resizebox{7cm}{!}{
  \begin{tabular}{l l }
    \toprule
    Methods& Computational Complexity  \\
    \midrule
    CACDR&   $\mathcal{O}(n*L*k*T)$ \\
   
    LFACDR & $\mathcal{O}((n+m)*L*k*T)$ \\
   
    DARec & $\mathcal{O}((n+m)*L*k*T)$   \\ 
  
    DDTCDR & $\mathcal{O}((n+m)*L*k*T)$     \\

    EMCDR & $\mathcal{O}((n+m)*k*T)$     \\   
    CST & $\mathcal{O}((n+m)*k*T)$     \\       

    LFM & $\mathcal{O}((n+m)*k*T)$     \\   
    \bottomrule
    
  \end{tabular}}
  \label{tab:complexity}
\end{table}

\begin{table*}[!t]
   \caption{The impact of different latent dimensions on Recommendation performance of the proposed models using the Movielens as source domain and the Netlix as target domain}
   \resizebox{\linewidth}{!}{
    \begin{tabular}{c||c|c||c|c||c| c|| c| c|| c| c}
    \hline
         Latent Dimension& \multicolumn{2}{c||}{k = 8} & \multicolumn{2}{c||}{k=32}  & \multicolumn{2}{c||}{k=64}  & \multicolumn{2}{c||}{k=128}  & \multicolumn{2}{c}{k=256} \\
         \cline{2-11} 
          & RMSE & MAE & RMSE & MAE & RMSE & MAE& RMSE & MAE& RMSE & MAE\\
        \hline
        CACDR & 0.1784 & 0.1391 & 0.1781 & 0.1389& 0.1738 & 0.1356& 0.1792 & 0.1400& 0.1806 & 0.1413 \\
        \hline
        LFACDR & 0.1747 & 0.1367 & 0.1739 &0.1359& 0.1733 & 0.1354& \textbf{0.1711} & \textbf{0.1331}& 0.1742 & 0.1361 \\
        \hline
    \end{tabular}}
    
    \label{tab:latent_dimension}
\end{table*}

\begin{table*}
  \caption{User-level relevance recommendation performance using the  Douban dataset}
  \resizebox{\linewidth}{!}{
  \begin{tabular}{ccccccccccl}
    \toprule
    Data&Metrics & CST & LFM & MF\_EMCDR & BPR\_EMCDR& DDTCDR & DARec& CACDR & LFACDR \\
    \midrule
    Book to Movie&RMSE & 0.2387 & 0.2347 & 0.2286 & 0.2245 & 0.1984& 0.1888  & 0.1652 & \textbf{0.1647} \\
             No.5    &MAE  & 0.1967& 0.1935 & 0.1863 & 0.1824 & 0.1587& 0.1506  & 0.1299 & \textbf{0.1276} \\
    \midrule
    Music to Book &RMSE & 0.2573 & 0.2420 & 0.2348 & 0.2311 & 0.2002& 0.1893  & 0.1751 & \textbf{0.1693} \\
            No.6     &MAE  & 0.2128& 0.2014 & 0.1987 & 0.1941 & 0.1597& 0.1461  & 0.1343 & \textbf{0.1313} \\

    \bottomrule
  \end{tabular}}
    \label{tab:results_douban}
\end{table*}

\begin{table}
\small
  \caption{
  {The performance  of the proposed models with and without Coupled learning stage using the  Douban Dataset} }
\centering
  {
  \begin{tabular}{cccccl}
    \toprule
      Data& Metrics  & CACDR& CACDR& LFACDR& LFACDR\\
      &   &  without & with &  without &  with \\
    \midrule
    No.5 &RMSE & 0.1730  & 0.1652  &  0.1728 & 0.1647  \\
        &MAE   & 0.1347 & 0.1299  &  0.1342 & 0.1276   \\
    \midrule
    N0.6&RMSE & 0.1836  & 0.1751&0.1785  & 0.1693  \\
        &MAE   & 0.1435  & 0.1343 &0.1411  & 0.1313   \\
    \bottomrule
  \end{tabular}}
    \label{tab:coupled_duban}
\end{table}

\begin{table}
\small
  \caption{
  {The performance  of the proposed models with and without Initialization stage using the  Douban dataset} }
\centering
  {
  \begin{tabular}{cccccl}
    \toprule
      Data& Metrics  & CACDR  & CACDR   & LFACDR   & LFACDR    \\
      &   &  without & with &  without &  with \\
    \midrule
    No.5&RMSE & 0.1725  & 0.1652  & 0.1746  & 0.1647  \\
        &MAE   & 0.1341 & 0.1299  & 0.1357  & 0.1276   \\
    \midrule
    No.6 &RMSE & 0.1812 & 0.1751& 0.1799 & 0.1693  \\
        &MAE   & 0.1421  & 0.1343 & 0.1412 & 0.1313   \\
    \bottomrule
  \end{tabular}}
    \label{tab:init_duban}
\end{table}
\section{Conclusions}
\label{sec:conc}

We have explored an item-level relevance CDR task where the source and the target domain contain common items  without sharing any additional information regarding the users' behavior, and thus avoiding the leak of user privacy. Additionally, we examined a user-level relevance scenario where the two related domains contain common users. We proposed  two novel coupled autoencoder-based deep learning methods for CDR that are able to represent the items in the source and target domains along with their coupled mapping function to model the non-linear relationships between these representations. The second method seeks to model the user and item-latent factors, while the first one does not make this assumption. 
We demonstrated some very promising results, in comparison to some popular methods on cross-domain recommendation. We used portions of the MovieLens, Netflix and Douban datasets with different sparsity levels and quantified the effect on our results. We also demonstrated the effect of learning the mapping function from one domain to the other, which turns out to be a significant part of the proposed method.

\section*{Acknowledgments}

This  work  is  partially  supported  by  the  Greek  Secretariat  for  Research  and  Innovation  and  the  EU, Project  Muselearn, T1EDK-00502  within  the  framework  of  “Competitiveness, Entrepreneurship and Innovation” (EPAnEK) Operational Programme 2014-2020.

\bibliographystyle{IEEEtran}
\bibliography{mybibfile}

\begin{thebibliography}{10}
\providecommand{\url}[1]{#1}
\csname url@samestyle\endcsname
\providecommand{\newblock}{\relax}
\providecommand{\bibinfo}[2]{#2}
\providecommand{\BIBentrySTDinterwordspacing}{\spaceskip=0pt\relax}
\providecommand{\BIBentryALTinterwordstretchfactor}{4}
\providecommand{\BIBentryALTinterwordspacing}{\spaceskip=\fontdimen2\font plus
\BIBentryALTinterwordstretchfactor\fontdimen3\font minus
  \fontdimen4\font\relax}
\providecommand{\BIBforeignlanguage}[2]{{%
\expandafter\ifx\csname l@#1\endcsname\relax
\typeout{** WARNING: IEEEtran.bst: No hyphenation pattern has been}%
\typeout{** loaded for the language `#1'. Using the pattern for}%
\typeout{** the default language instead.}%
\else
\language=\csname l@#1\endcsname
\fi
#2}}
\providecommand{\BIBdecl}{\relax}
\BIBdecl

\bibitem{RCR}
\BIBentryALTinterwordspacing
D.~H. Park, H.~K. Kim, I.~Y. Choi, and J.~K. Kim, ``A literature review and
  classification of recommender systems research,'' \emph{Expert Systems with
  Applications}, vol.~39, no.~11, pp. 10\,059--10\,072, 2012. [Online].
  Available:
  \url{https://www.sciencedirect.com/science/article/pii/S0957417412002825}
\BIBentrySTDinterwordspacing

\bibitem{10.1016/j.neucom.2016.12.102}
\BIBentryALTinterwordspacing
Q.-Y. Hu, Z.-L. Zhao, C.-D. Wang, and J.-H. Lai, ``An item orientated
  recommendation algorithm from the multi-view perspective,''
  \emph{Neurocomput.}, vol. 269, no.~C, p. 261–272, Dec. 2017. [Online].
  Available: \url{https://doi.org/10.1016/j.neucom.2016.12.102}
\BIBentrySTDinterwordspacing

\bibitem{Ma2019}
\BIBentryALTinterwordspacing
J.~Ma, J.~Wen, M.~Zhong, W.~Chen, and X.~Li, ``{MMM: Multi-source Multi-net
  Micro-video Recommendation with Clustered Hidden Item Representation
  Learning},'' \emph{Data Science and Engineering 2019 4:3}, vol.~4, no.~3, pp.
  240--253, sep 2019. [Online]. Available:
  \url{https://link.springer.com/article/10.1007/s41019-019-00101-4}
\BIBentrySTDinterwordspacing

\bibitem{Bobadilla2013}
J.~Bobadilla, F.~Ortega, A.~Hernando, and A.~Guti{\'{e}}rrez, ``{Recommender
  systems survey},'' \emph{Knowledge-Based Systems}, vol.~46, pp. 109--132, jul
  2013.

\bibitem{cold}
C.~Wang and D.~M. Blei, ``Collaborative topic modeling for recommending
  scientific articles,'' in \emph{Proceedings of the 17th ACM SIGKDD
  International Conference on Knowledge Discovery and Data Mining}, ser. KDD
  '11.\hskip 1em plus 0.5em minus 0.4em\relax Association for Computing
  Machinery, 2011, p. 448–456.

\bibitem{Natarajan2020}
S.~Natarajan, S.~Vairavasundaram, S.~Natarajan, and A.~H. Gandomi, ``{Resolving
  data sparsity and cold start problem in collaborative filtering recommender
  system using Linked Open Data},'' \emph{Expert Systems with Applications},
  vol. 149, p. 113248, jul 2020.

\bibitem{Zhang20201}
\BIBentryALTinterwordspacing
Z.~Zhang, Y.~Zhang, and Y.~Ren, ``{Employing neighborhood reduction for
  alleviating sparsity and cold start problems in user-based collaborative
  filtering},'' \emph{Information Retrieval Journal 2020 23:4}, vol.~23, no.~4,
  pp. 449--472, jun 2020. [Online]. Available:
  \url{https://link.springer.com/article/10.1007/s10791-020-09378-w}
\BIBentrySTDinterwordspacing

\bibitem{cross_1}
A.~M. Elkahky, Y.~Song, and X.~He, ``A multi-view deep learning approach for
  cross domain user modeling in recommendation systems,'' in \emph{Proceedings
  of the 24th International Conference on World Wide Web}, ser. WWW '15.\hskip
  1em plus 0.5em minus 0.4em\relax International World Wide Web Conferences
  Steering Committee, 2015, p. 278–288.

\bibitem{app1}
\BIBentryALTinterwordspacing
D.~Cao, X.~He, L.~Nie, X.~Wei, X.~Hu, S.~Wu, and T.-S. Chua, ``Cross-platform
  app recommendation by jointly modeling ratings and texts,'' \emph{ACM Trans.
  Inf. Syst.}, vol.~35, no.~4, Jul. 2017. [Online]. Available:
  \url{https://doi.org/10.1145/3017429}
\BIBentrySTDinterwordspacing

\bibitem{app2}
\BIBentryALTinterwordspacing
T.-H. Lin, C.~Gao, and Y.~Li, ``Cross: Cross-platform recommendation for social
  e-commerce,'' in \emph{Proceedings of the 42nd International ACM SIGIR
  Conference on Research and Development in Information Retrieval}, ser.
  SIGIR'19.\hskip 1em plus 0.5em minus 0.4em\relax New York, NY, USA:
  Association for Computing Machinery, 2019, p. 515–524. [Online]. Available:
  \url{https://doi.org/10.1145/3331184.3331191}
\BIBentrySTDinterwordspacing

\bibitem{app3}
\BIBentryALTinterwordspacing
T.~Mei, B.~Yang, X.-S. Hua, and S.~Li, ``Contextual video recommendation by
  multimodal relevance and user feedback,'' \emph{ACM Trans. Inf. Syst.},
  vol.~29, no.~2, Apr. 2011. [Online]. Available:
  \url{https://doi.org/10.1145/1961209.1961213}
\BIBentrySTDinterwordspacing

\bibitem{app4}
\BIBentryALTinterwordspacing
J.~Tang, S.~Wu, J.~Sun, and H.~Su, ``Cross-domain collaboration
  recommendation,'' in \emph{Proceedings of the 18th ACM SIGKDD International
  Conference on Knowledge Discovery and Data Mining}, ser. KDD '12.\hskip 1em
  plus 0.5em minus 0.4em\relax New York, NY, USA: Association for Computing
  Machinery, 2012, p. 1285–1293. [Online]. Available:
  \url{https://doi.org/10.1145/2339530.2339730}
\BIBentrySTDinterwordspacing

\bibitem{8392508}
C.-D. Wang, Z.-H. Deng, J.-H. Lai, and P.~S. Yu, ``Serendipitous recommendation
  in e-commerce using innovator-based collaborative filtering,'' \emph{IEEE
  Transactions on Cybernetics}, vol.~49, no.~7, pp. 2678--2692, 2019.

\bibitem{pmlr-v38-iwata15}
T.~Iwata and T.~Koh, ``{Cross-domain recommendation without shared users or
  items by sharing latent vector distributions},'' in \emph{Proceedings of the
  18th International Conference on Artificial Intelligence and Statistics},
  vol.~38.\hskip 1em plus 0.5em minus 0.4em\relax PMLR, 09--12 May 2015, pp.
  379--387.

\bibitem{Cantador2015}
I.~Cantador, I.~Fern{\'{a}}ndez-Tob{\'{i}}as, S.~Berkovsky, and P.~Cremonesi,
  ``{Cross-domain recommender systems},'' in \emph{Recommender Systems
  Handbook, Second Edition}.\hskip 1em plus 0.5em minus 0.4em\relax Springer
  US, jan 2015, pp. 919--959.

\bibitem{Zhu2021}
F.~Zhu, Y.~Wang, C.~Chen, J.~Zhou, L.~Li, and G.~Liu, ``Cross-domain
  recommendation: Challenges, progress, and prospects,'' in \emph{IJCAI}, 2021.

\bibitem{Zhong2020}
S.-T. Zhong, L.~Huang, C.-D. Wang, J.-H. Lai, and P.~S. Yu, ``An autoencoder
  framework with attention mechanism for cross-domain recommendation,''
  \emph{IEEE Transactions on Cybernetics}, pp. 1--13, 2020.

\bibitem{Moreno2012}
O.~Moreno, B.~Shapira, L.~Rokach, and G.~Shani, ``{TALMUD: Transfer learning
  for multiple domains},'' in \emph{ACM International Conference Proceeding
  Series}.\hskip 1em plus 0.5em minus 0.4em\relax ACM Press, 2012, pp.
  425--434.

\bibitem{text1}
X.~Xin, Z.~Liu, C.-Y. Lin, H.~Huang, X.~Wei, and P.~Guo, ``Cross-domain
  collaborative filtering with review text,'' in \emph{Proceedings of
  International Joint Conference on Artificial Intelligence}, 2015, pp.
  1827--1834.

\bibitem{text2}
W.~Fu, Z.~Peng, S.~Wang, Y.~Xu, and J.~Li, ``{Deeply fusing reviews and
  contents for cold start users in cross-domain recommendation systems},'' in
  \emph{33rd AAAI Conference on Artificial Intelligence}, vol.~33,
  no.~01.\hskip 1em plus 0.5em minus 0.4em\relax AAAI Press, jul 2019, pp.
  94--101.

\bibitem{Gao2019}
C.~Gao, K.~Zhao, X.~Chen, X.~He, D.~Jin, F.~Feng, and Y.~Li, ``{Cross-domain
  recommendation without sharing user-relevant data},'' in \emph{The World Wide
  Web Conference}.\hskip 1em plus 0.5em minus 0.4em\relax Association for
  Computing Machinery, Inc, may 2019, pp. 491--502.

\bibitem{He2019}
M.~He, J.~Zhang, and S.~Zhang, ``{ACTL: Adaptive Codebook Transfer Learning for
  Cross-Domain Recommendation},'' \emph{IEEE Access}, vol.~7, pp.
  19\,539--19\,549, 2019.

\bibitem{10.1145/3366423.3380036}
\BIBentryALTinterwordspacing
J.~Liu, P.~Zhao, F.~Zhuang, Y.~Liu, V.~S. Sheng, J.~Xu, X.~Zhou, and H.~Xiong,
  ``Exploiting aesthetic preference in deep cross networks for cross-domain
  recommendation,'' in \emph{Proceedings of The Web Conference 2020}, ser. WWW
  '20.\hskip 1em plus 0.5em minus 0.4em\relax New York, NY, USA: Association
  for Computing Machinery, 2020, p. 2768–2774. [Online]. Available:
  \url{https://doi.org/10.1145/3366423.3380036}
\BIBentrySTDinterwordspacing

\bibitem{man}
T.~Man, H.~Shen, X.~Jin, and X.~Cheng, ``Cross-domain recommendation: An
  embedding and mapping approach,'' in \emph{Proceedings of International Joint
  Conference on Artificial Intelligence, {IJCAI-17}}, 2017, pp. 2464--2470.

\bibitem{zhang}
Z.~Zhang, X.~Jin, L.~Li, G.~Ding, and Q.~Yang, ``Multi-domain active learning
  for recommendation,'' in \emph{Proceedings of the Thirtieth AAAI Conference
  on Artificial Intelligence}, ser. AAAI'16.\hskip 1em plus 0.5em minus
  0.4em\relax AAAI Press, 2016, p. 2358–2364.

\bibitem{T1}
J.~Shang, M.~Sun, and K.~Collins-Thompson, ``Demographic inference via
  knowledge transfer in cross-domain recommender systems,'' in \emph{2018 IEEE
  International Conference on Data Mining (ICDM)}, 2018, pp. 1218--1223.

\bibitem{T2}
J.~Manotumruksa, D.~Rafailidis, C.~Macdonald, and I.~Ounis, ``On cross-domain
  transfer in venue recommendation,'' in \emph{Advances in Information
  Retrieval}, L.~Azzopardi, B.~Stein, N.~Fuhr, P.~Mayr, C.~Hauff, and
  D.~Hiemstra, Eds.\hskip 1em plus 0.5em minus 0.4em\relax Cham: Springer
  International Publishing, 2019, pp. 443--456.

\bibitem{BLin}
B.~Li, Q.~Yang, and X.~Xue, ``Can movies and books collaborate? cross-domain
  collaborative filtering for sparsity reduction,'' in \emph{Proceedings of the
  21st International Joint Conference on Artificial Intelligence}, ser.
  IJCAI'09, 2009, p. 2052–2057.

\bibitem{Loni}
B.~Loni, Y.~Shi, M.~Larson, and A.~Hanjalic, ``Cross-domain collaborative
  filtering with factorization machines,'' in \emph{Advances in Information
  Retrieval}, M.~de~Rijke, T.~Kenter, A.~P. de~Vries, C.~Zhai, F.~de~Jong,
  K.~Radinsky, and K.~Hofmann, Eds.\hskip 1em plus 0.5em minus 0.4em\relax
  Cham: Springer International Publishing, 2014, pp. 656--661.

\bibitem{Darec}
F.~Yuan, L.~Yao, and B.~Benatallah, ``Darec: Deep domain adaptation for
  cross-domain recommendation via transferring rating patterns,'' in
  \emph{Proceedings of the 28th International Joint Conference on Artificial
  Intelligence}, ser. IJCAI'19.\hskip 1em plus 0.5em minus 0.4em\relax AAAI
  Press, 2019, p. 4227–4233.

\bibitem{Singh2008}
A.~P. Singh and G.~J. Gordon, ``{Relational learning via collective matrix
  factorization},'' in \emph{Proceedings of the ACM SIGKDD International
  Conference on Knowledge Discovery and Data Mining}.\hskip 1em plus 0.5em
  minus 0.4em\relax ACM Press, 2008, pp. 650--658.

\bibitem{Pan2010}
W.~Pan, E.~W. Xiang, N.~N. Liu, and Q.~Yang, ``{Transfer Learning in
  Collaborative Filtering for Sparsity Reduction},'' in \emph{Proceedings of
  the AAAI Conference on Artificial Intelligence}, vol.~24, no.~1, jul 2010.

\bibitem{lfm}
D.~Agarwal, B.-C. Chen, and B.~Long, ``Localized factor models for
  multi-context recommendation,'' in \emph{Proceedings of the 17th ACM SIGKDD
  International Conference on Knowledge Discovery and Data Mining}.\hskip 1em
  plus 0.5em minus 0.4em\relax Association for Computing Machinery, 2011, p.
  609–617.

\bibitem{CCCFNet}
\BIBentryALTinterwordspacing
J.~Lian, F.~Zhang, X.~Xie, and G.~Sun, ``Cccfnet: A content-boosted
  collaborative filtering neural network for cross domain recommender
  systems,'' in \emph{Proceedings of the 26th International Conference on World
  Wide Web Companion}, ser. WWW '17 Companion.\hskip 1em plus 0.5em minus
  0.4em\relax Republic and Canton of Geneva, CHE: International World Wide Web
  Conferences Steering Committee, 2017, p. 817–818. [Online]. Available:
  \url{https://doi.org/10.1145/3041021.3054207}
\BIBentrySTDinterwordspacing

\bibitem{cluster1}
S.~Gao, H.~Luo, D.~Chen, S.~Li, P.~Gallinari, and J.~Guo, ``Cross-domain
  recommendation via cluster-level latent factor model,'' in \emph{Machine
  Learning and Knowledge Discovery in Databases}, H.~Blockeel, K.~Kersting,
  S.~Nijssen, and F.~{\v{Z}}elezn{\'y}, Eds.\hskip 1em plus 0.5em minus
  0.4em\relax Berlin, Heidelberg: Springer Berlin Heidelberg, 2013, pp.
  161--176.

\bibitem{cluster2}
\BIBentryALTinterwordspacing
N.~Mirbakhsh and C.~X. Ling, ``Improving top-n recommendation for cold-start
  users via cross-domain information,'' \emph{ACM Trans. Knowl. Discov. Data},
  vol.~9, no.~4, Jun. 2015. [Online]. Available:
  \url{https://doi.org/10.1145/2724720}
\BIBentrySTDinterwordspacing

\bibitem{rafa}
\BIBentryALTinterwordspacing
D.~Rafailidis and F.~Crestani, ``A collaborative ranking model for cross-domain
  recommendations,'' in \emph{Proceedings of the 2017 ACM on Conference on
  Information and Knowledge Management}, ser. CIKM '17.\hskip 1em plus 0.5em
  minus 0.4em\relax New York, NY, USA: Association for Computing Machinery,
  2017, p. 2263–2266. [Online]. Available:
  \url{https://doi.org/10.1145/3132847.3133107}
\BIBentrySTDinterwordspacing

\bibitem{cluster3}
\BIBentryALTinterwordspacing
Y.~Wang, C.~Feng, C.~Guo, Y.~Chu, and J.-N. Hwang, ``Solving the sparsity
  problem in recommendations via cross-domain item embedding based on
  co-clustering,'' in \emph{Proceedings of the Twelfth ACM International
  Conference on Web Search and Data Mining}, ser. WSDM '19.\hskip 1em plus
  0.5em minus 0.4em\relax New York, NY, USA: Association for Computing
  Machinery, 2019, p. 717–725. [Online]. Available:
  \url{https://doi.org/10.1145/3289600.3290973}
\BIBentrySTDinterwordspacing

\bibitem{Elkahky}
\BIBentryALTinterwordspacing
A.~M. Elkahky, Y.~Song, and X.~He, ``A multi-view deep learning approach for
  cross domain user modeling in recommendation systems,'' in \emph{Proceedings
  of the 24th International Conference on World Wide Web}, ser. WWW '15.\hskip
  1em plus 0.5em minus 0.4em\relax Republic and Canton of Geneva, CHE:
  International World Wide Web Conferences Steering Committee, 2015, p.
  278–288. [Online]. Available: \url{https://doi.org/10.1145/2736277.2741667}
\BIBentrySTDinterwordspacing

\bibitem{At}
F.~Zhu, Y.~Wang, C.~Chen, G.~Liu, and X.~Zheng, ``A graphical and attentional
  framework for dual-target cross-domain recommendation,'' in \emph{Proceedings
  of the Twenty-Ninth International Joint Conference on Artificial
  Intelligence, {IJCAI-20}}, C.~Bessiere, Ed.\hskip 1em plus 0.5em minus
  0.4em\relax International Joint Conferences on Artificial Intelligence
  Organization, 7 2020, pp. 3001--3008, main track.

\bibitem{Kanagawa}
H.~Kanagawa, H.~Kobayashi, N.~Shimizu, Y.~Tagami, and T.~Suzuki, ``Cross-domain
  recommendation via deep domain adaptation,'' in \emph{Advances in Information
  Retrieval}, L.~Azzopardi, B.~Stein, N.~Fuhr, P.~Mayr, C.~Hauff, and
  D.~Hiemstra, Eds.\hskip 1em plus 0.5em minus 0.4em\relax Cham: Springer
  International Publishing, 2019, pp. 20--29.

\bibitem{Zhao1}
\BIBentryALTinterwordspacing
C.~Zhao, C.~Li, and C.~Fu, ``Cross-domain recommendation via preference
  propagation graphnet,'' in \emph{Proceedings of the 28th ACM International
  Conference on Information and Knowledge Management}, ser. CIKM '19.\hskip 1em
  plus 0.5em minus 0.4em\relax New York, NY, USA: Association for Computing
  Machinery, 2019, p. 2165–2168. [Online]. Available:
  \url{https://doi.org/10.1145/3357384.3358166}
\BIBentrySTDinterwordspacing

\bibitem{Ma2021}
G.~Ma, Y.~Wang, X.~Zheng, X.~Miao, and Q.~Liang, ``{A trust-aware latent space
  mapping approach for cross-domain recommendation},'' \emph{Neurocomputing},
  vol. 431, pp. 100--110, mar 2021.

\bibitem{Conet}
\BIBentryALTinterwordspacing
G.~Hu, Y.~Zhang, and Q.~Yang, ``Conet: Collaborative cross networks for
  cross-domain recommendation,'' in \emph{Proceedings of the 27th ACM
  International Conference on Information and Knowledge Management}, ser. CIKM
  '18.\hskip 1em plus 0.5em minus 0.4em\relax New York, NY, USA: Association
  for Computing Machinery, 2018, p. 667–676. [Online]. Available:
  \url{https://doi.org/10.1145/3269206.3271684}
\BIBentrySTDinterwordspacing

\bibitem{Li2020}
P.~Li and A.~Tuzhilin, ``{DDTCDR: Deep dual transfer cross domain
  recommendation},'' \emph{WSDM 2020 - Proceedings of the 13th International
  Conference on Web Search and Data Mining}, pp. 331--339, jan 2020.

\bibitem{9640532}
Q.~Zhang, W.~Liao, G.~Zhang, B.~Yuan, and J.~Lu, ``A deep dual adversarial
  network for cross-domain recommendation,'' \emph{IEEE Transactions on
  Knowledge and Data Engineering}, pp. 1--1, 2021.

\bibitem{8698453}
C.~Wang, M.~Niepert, and H.~Li, ``Recsys-dan: Discriminative adversarial
  networks for cross-domain recommender systems,'' \emph{IEEE Transactions on
  Neural Networks and Learning Systems}, vol.~31, no.~8, pp. 2731--2740, 2020.

\bibitem{8525418}
Q.~Zhang, J.~Lu, D.~Wu, and G.~Zhang, ``A cross-domain recommender system with
  kernel-induced knowledge transfer for overlapping entities,'' \emph{IEEE
  Transactions on Neural Networks and Learning Systems}, vol.~30, no.~7, pp.
  1998--2012, 2019.

\bibitem{ctan}
\BIBentryALTinterwordspacing
C.~Zhao, C.~Li, R.~Xiao, H.~Deng, and A.~Sun, ``Catn: Cross-domain
  recommendation for cold-start users via aspect transfer network,'' in
  \emph{Proceedings of the 43rd International ACM SIGIR Conference on Research
  and Development in Information Retrieval}, ser. SIGIR '20.\hskip 1em plus
  0.5em minus 0.4em\relax New York, NY, USA: Association for Computing
  Machinery, 2020, p. 229–238. [Online]. Available:
  \url{https://doi.org/10.1145/3397271.3401169}
\BIBentrySTDinterwordspacing

\bibitem{da}
S.~J. Pan and Q.~Yang, ``A survey on transfer learning,'' \emph{IEEE
  Transactions on Knowledge and Data Engineering}, vol.~22, no.~10, pp.
  1345--1359, 2010.

\bibitem{Ca2010}
P.~V. Ca, L.~T. Edu, I.~Lajoie, Y.~B. Ca, and P.-A.~M. Ca, ``{Stacked Denoising
  Autoencoders: Learning Useful Representations in a Deep Network with a Local
  Denoising Criterion Pascal Vincent Hugo Larochelle Yoshua Bengio
  Pierre-Antoine Manzagol},'' Tech. Rep., 2010.

\bibitem{Kuchaiev2017}
O.~Kuchaiev and B.~Ginsburg, ``{Training Deep AutoEncoders for Collaborative
  Filtering},'' \emph{arXiv}, aug 2017.

\end{thebibliography}


\newpage

 




\vfill

\end{document}